\documentclass[12pt]{article}
\usepackage{fullpage, amsmath, amsthm, amsfonts}
\usepackage{hyperref}
\usepackage{nameref}
\usepackage{lmodern}
\usepackage{algorithm}
\usepackage{algorithmicx}
\usepackage{color}
\usepackage{algpseudocode}
\usepackage{graphicx}
\usepackage{afterpage}
\usepackage{parskip}
\usepackage{natbib}           
\setcitestyle{aysep={}}
\usepackage{setspace}
\usepackage{titlesec}

\newcommand{\btheta}{\boldsymbol{\theta}}

\newcommand{\bx}{\mathbf{x}}

\newcommand{\bA}{\mathbf{A}}

\newcommand{\bc}{\mathbf{c}}

\newcommand{\boldf}{\mathbf{f}}

\newcommand{\bq}{\mathbf{q}}
\newcommand{\bz}{\mathbf{z}}

\newcommand{\bbeta}{\boldsymbol{\beta}}

\newcommand{\bSigma}{\boldsymbol{\Sigma}}
\newcommand{\bB}{\mathbf{B}}
\newcommand{\bD}{\mathbf{D}}
\newcommand{\bpsi}{\psi}

\title{A Probabilistic Model for Analyzing Summary Birth History Data}
\author{Katie Wilson$^1$ and Jon Wakefield$^{1,2}$\\
\small{$^1$ Department of Biostatistics, University of Washington} \\
\small{$^2$ Department of Statistics, University of Washington}}
\date{}
\begin{document}
\clearpage\maketitle
\thispagestyle{empty}


\newpage

\pagestyle{empty} 

\begin{large}{\bf Abstract} \end{large}

{\bf BACKGROUND}

There is an increasing demand for high quality subnational estimates of under-five mortality. 
In low and middle income countries, where the burden of under-five mortality is concentrated, vital registration is often lacking and household surveys, which provide full birth history  data, are often the most reliable source. Unfortunately, these data are spatially sparse and so data are pulled from other sources to increase the available information. Summary birth histories represent a large fraction of the available data, and provide numbers of births and deaths aggregated over time, along with the mother's age. 

{\bf OBJECTIVE}

Specialized methods are needed to leverage this information, and previously the Brass method, and variants, have been used. We wish to develop a model-based approach that can propagate errors, and make the most efficient use of the data. Further, we strive to provide a method that does not have large computational overhead.

{\bf CONTRIBUTION} 

We describe a computationally efficient model-based approach  which allows summary birth history and full birth history data to be combined into analyses of under-five mortality in a natural way. The method is based on fertility and mortality models that allow direct smoothing over time and space, with the possibility for including relevant covariates that are associated with fertility and/or mortality. We first examine the behavior of the approach on simulated data, before applying the model to survey and census data from Malawi.


\clearpage

\section{Introduction}

The United Nations' Sustainable Developmental Goals emphasize the importance of subnational estimates of the under-five mortality rate (U5MR). Subnational estimates of U5MR are an important measure of the health of a nation and are used to inform public policy. Unfortunately, countries where U5MR is highest often lack vital registration systems that track births and deaths. Instead, data is collected from household surveys and censuses. Many surveys, such as the Demographic and Health Surveys (DHS), provide full birth history (FBH) data, which consists of the dates of birth and death (if applicable) of the children of surveyed women. FBH data can be used to obtain time-varying subnational estimates of U5MR (in regions where enough FBH data is collected), see for example \cite{li:etal:19}. Many censuses (and some surveys) collect summary birth history (SBH) data in which the available birth history data is the number of births and deaths of surveyed women. SBH information is far easier to collect and hence its inclusion on the census in many low and middle income countries (LMIC). No temporal information on when the births and deaths occurred is provided in SBH data; therefore, inclusion of this data is more challenging. However, the SBH data benefits from the large sample size of the census. With both FBH and SBH data, other variables, such as the age of the woman, are also available.  SBH data is more common; in a study of U5MR in Africa, 90\% of the births were from SBH data \citep{golding:etal:17}.

The wish to exploit SBH data has long been desired, beginning with the famous approach introduced by \cite{brass:64}. The Brass method, which we describe in detail in the next section, is essentially a deterministic approach, which has seen many variants since its introduction
 \citep{brass:75,coale:trussell:77,feeney:76,manualX:83,sullivan:72,trussell:75}. A commonly used approach for obtaining standard errors for Brass estimates, which may be used to construct confidence intervals, is the jackknife \citep{pedersenandliu:2012}.  \cite{rajaratnam:etal:10} refine this method and use observed FBH data in place of model life tables. More recently, \cite{burstein:etal:18}  describe an extension of this approach which  introduced a discrete survival model, with birth times assigned to SBH children. \cite{verhulst:16} compares various SBH models using simulated data.  \cite{hill:etal:15} propose two alternatives, including the Birth History Imputation method, in which SBH women are matched to FBH women who are of approximately the same age, and have the same numbers of both births and  deaths. They also describe the Cohort Change method, which requires data from SBH surveys taken 1--2 years apart. They posit that the change in number of children that died and number of children born will largely be driven by U5MR and thus leverage these observed quantities in the SBH data to derive an estimate of U5MR. Unfortunately, the two methods provided very mixed results when applied to real data \citep{brady:hill:17}.
 
 Perhaps surprisingly, the above body of work for the analysis of SBH data do not take a fully probabilistic approach in which a model is specified at the child level, with uncertainty correctly propagated. 
 By ``fully probabilistic", we mean a model in which probability distributions are specified for both the birth and death times. 
 Such a fully stochastic model allows FBH and SBH data to be combined in a coherent way, since they are both built on common underlying  models for fertility and death. The approach can be tuned to the context at hand; for example, spatial and temporal smoothers can be  incorporated, along with indicators of urban/rural and other covariates. Further, bias terms can be included to allow for systematic differences that may exist in particular data sources. Inference follows in a straightforward fashion, if a regular statistical modeling framework is used. 

In the context of analyzing FBH data, a modeling framework based on space-time modeling of discrete hazards is available, and summarized in \cite{wakefield:etal:19}. Previously, \cite{wilson:wakefield:20SBH} proposed a model-based approach in which the unknown birth and death times in SBH data are simulated and then combined with FBH data to obtain estimates of U5MR in a data augmentation framework. While this method allows for flexible inclusion of covariates, and can accommodate bias in surveys, it is computationally demanding, since typically the births and deaths times of millions of children must be repeatedly simulated. Unfortunately, prior to  \cite{wilson:wakefield:20SBH},  approaches to analyzing SBH data did not support a space-time modeling  framework, since they were not based on providing probability models directly for the mortality hazards. 

In this paper, we propose a computationally efficient approach using a Poisson approximation that removes the simulation step. By greatly easing the computational burden we can extend the modeling approach and incorporate more complex space-time models.  We fit the model using the \texttt{R} package {\tt TMB} (with the acronym standing for Template Model Builder), which allows one to quickly obtain empirical Bayes estimates along with other inferential summaries \citep{kristensen:14,kristensen:etal:16}. The paper is laid out as follows. We begin by describing the Brass method, before describing the proposed method. We describe a simulation study, in which we study the performance of our method and then apply the method to data from Malawi. FBH data from five households surveys is combined with census SBH data to give yearly U5MR estimates in 26 regions of Malawi; this is the first comprehensive analysis of child mortality data  in Malawi that rigorously combines FBH and SBH data. The paper concludes with a discussion. 


\section{Methods}

\subsection{The Brass Method}

We describe how to implement the Trussell version of the Brass method, which uses the Coale-Demeny model life tables, this description is adapted from Chapter III of \cite{manualX:83}, using the same notation. We index the 5-year age groups of women, 15--19,\dots,45--49, by $i=1,\dots, 7$, and let $q(x)$ represent the probability that a child dies before age $x$. The basic idea is to equate the fractions of children who died to mothers of different ages, to mortality probabilities $q(x)$, for an appropriate $x$.
The method is described by the following steps: 
\begin{itemize}
\item[(1)] Calculate the average parity per woman by age group $i$, $$P(i) = \frac{CEB(i)}{{FP}(i)}$$
where $CEB(i)$ are the number of children ever born to, and ${FP}(i)$ are the total number of women in age group $i$.
\item[(2)] Calculate the fraction of children dead in mothers age group $i$, denoted $D(i)$, with
$$D(i) = \frac{CD(i)}{CEB(i)},$$
where $CD(i)$ is the number of children who died to mothers in age group $i$.
\item[(3)] Next, convert the fraction of children dead in mothers age group $i$ to $q(x)$ for a particular $x$. Specifically, calculate the probability of dying by age $x$,
$$q(x) = D(i) \left(a(i) + b(i) \frac{P(1)}{P(2)} + c(i) \frac{P(2)}{P(3)}\right)$$
where the coefficients $a(i),b(i),c(i)$  are estimated from a simulation based on the Coale-Demeny model life tables and $x$ is given in the table in the Supplementary Materials. For example, the proportions of deaths in mothers age group 15--19 maps to $q(1)$ and those in the 30--34 group to $q(5)$.
\item[(4)] Given that mortality and fertility change over time,  appropriate reference dates for $q(x)$ are calculated as,
$$t(i) = e(i) + f(i) \frac{P(1)}{P(2)} + g(i) \frac{P(2)}{P(3)}$$
where the coefficients $e(i),f(i),g(i)$ are estimated from a simulation based on the Coale-Demeny model life tables.
\item[(5)] If U5MR is of interest, convert $q(x)$ to  $q(5)$. This is done by identifying the mortality ``level'' (expectation of life) in a table that most closely corresponds to the observed $q(x)$ and then using the table to convert $q(x)$ to the desired index. There are different tables for each of the Coale-Demeny models. In practice, linear interpolation is used to interpolate between different levels in the life table. Suppose $q^e(x)$ is the estimated value of $q(x)$ obtained in step 3. In the table, find the level $j$ such that
$$q^{j+1}(x) < q^e(x) < q^j(x).$$
Therefore,
\begin{align*}
q(5) & = (1-h) q^j(5) + hq^{j+1}(5)\\
h & =\frac{q^e(x)-q^j(x)}{q^{j+1}(x) - q^j(x)}.
\end{align*}
\end{itemize}
The Brass method has been widely used, and has formed the basis for many of the proposed methods for leveraging SBH data. The approach we describe next, follows a different tack, by beginning with probability models for fertility and mortality.

\subsection{
An Intuitive Derivation of the New Method}




The approach is based on a plausible full probability model, i.e.,~a model that could be used to simulate birth and death data. We then consider taking a survey, in which data on births and deaths is gathered, and describe a model for the data in the case when we have access to all birth and death times. We finally write down the model for the SBH observed data, in which we average over the unobserved data, and then approximate this form in order to obtain a computationally efficient model. We initially do this in the simplified case in which fertility and mortality do not change over time or with covariates. \cite{preston:etal:01} provide a clear derivation of the Brass method that is based on a mean value theorem.  The Brass approach provides  neither a probability model which can be leveraged for smoothing, nor an obvious way by which inference can be performed.
Below we show that, under well-defined  modeling assumptions and approximations, the Brass method can be seen as a method of moments.

 Let $Y_m=0/1$ be a binary indicator for a non-birth/birth when a woman is of age $m$, with $Y_m | f_m \sim \mbox{Bernoulli}(f_m)$ where $f_m$ is the probability of giving birth for a woman of age $m$, for $m=m_{\min},\dots,m_{\max}$. Following birth, we use a discrete time hazards model for death. For a child alive at age $a$, let $Z_a=0/1$ be the binary indicator of non-death/death before age $a+1$, given survival until age $a$. The distribution of the death indicators is
 $Z_a | _1q_a \sim \mbox{Bernoulli}(_1q_a)$ where $_1q_a$ is the (conditional) probability of death between ages $a$ and $a+1$, for $a=0,1,\dots$. Define $\bq$ to be the vector containing these conditional probabilities of death. Now suppose that a survey is taken and $M$ mothers of different ages are interviewed; we focus on data from mothers of age $m_s$. 
 Let $\bB^{m_s}=[B_0,\dots,B_{m_s}]$ and $\bD^{m_s}= [D_0,\dots,D_{m_s}]$ be the collections of births and deaths from these mothers. Here, $B_a$ is the number of births born to women $a$ years before the survey and $D_a$ are the number of those children who died before the survey. These variables would be available for FBH data. 
The total numbers of births and deaths are defined as $T_B^{m_s}=\sum_{a=0}^{m_s} B_a$ and $T_D^{m_s}=\sum_{a=0}^{m_s}  D_a$.  These latter are what constitute the SBH data, but to derive a probability model for the data, we first write down a model for the full data.

For  births $a$ years before the survey, we assume the distribution of the number of deaths is,
\begin{equation}\label{eq:obs1} D_a | B_a, \bq \sim  \mbox{Binomial}(B_a, q(a) ),\end{equation}
where $q(a) = 1 - \prod_{i=0}^{a-1}(1 -\, _1q_i)$ is the probability that a child born $a$ years before the survey dies. A  distribution for the collection of births is more difficult to write down (because the timings of births to the same mother are necessarily highly dependent)  but, as we will see, this will not be needed. Denote the births distribution by $\Pr ( \bB^{m_s} | T_B^{m_s},\bc)$ with 
$\bc=[c_0,\dots,c_{m_s}]$ and where $c_a = f_a/\sum_{a'=0}^{m_{s}} f_{a'}$ 
are the probabilities a woman gives birth $a$ years before the survey, given a birth occurs in the time interval $(m_0,m_{m_{s}})$. 
Given the total number of births, the mean number of birth $a$ year before the survey is, 
\begin{equation}\label{eq:obs2}E[ B_a | T_B^{m_s}, c_a] = T_B^{m_s} c_a. \end{equation}
Hence, we have forms upon which estimation of birth and death probabilities could be based, if we had complete data, as would be available in a FBH survey. To obtain a sampling model for the observed SBH data, we average over the missing information, first the unknown death times and then the unknown birth times.

When only the total number of deaths, $T_D^{m_s}$, is observed, the distribution of the sum of the deaths is a { convolution} of the distributions given in Eq.~(\ref{eq:obs1}). This is computationally intractable, since all possible legal combinations of deaths need to be enumerated, see \cite{wakefield:04read}. However, if we approximate the binomials by Poissons (which is valid if $q(a)$ is relatively small and the number $B_a$ is large), we obtain a distribution for the total number of deaths,
$$ T_D^{m_s} | \bB^{m_s}, \bq \sim \mbox{ Poisson } \left( \sum_{a=0}^{m_s} B_a q(a) \right).$$
Let $S_{m_s}$ be the set of legal configurations of births that can lead to a birth total of $T_B^{m_s}$. Averaging over this set of unknown births gives the { mixture distribution}:
$$T_D^{m_s} | T_B^{m_s} , \bc, \bq \sim \sum_{\bB^{{m_s}} \in S_{m_s}} \Pr ( \bB^{m_s} | T_B^{m_s}, \bc)\times  \mbox{ Poisson } \left( \sum_{a=0}^{m_s} B_a q(a) \right).$$
This expression looks complex, but it is quite intuitive, since the Poisson model describes the probability of the deaths, given a particular configuration of births and these Poissons are averaged over the uncertainty in when the births occur.
Unfortunately, this expression depends on the birth times to give the birth totals over time $ \bB^{m_s}$, and these are unobserved in the SBH data.

Rather than average over all elements in the sum, which is costly, we replace the distribution by the $m_s+1$ means of $ \Pr ( \bB^{m_s} | T_B^{m_s})$, which are given in Eq.~(\ref{eq:obs2}), to give
\begin{equation}\label{eq:obslik}
T_D^{m_s} | T_B^{m_s} , \bc, \bq \sim \mbox{ Poisson } \left( T_B^{m_s} \sum_{a=0}^{m_s} c_a q(a) \right).
\end{equation}

Hence, we are conditioning on an estimate of the expectation of the number of births, rather than averaging over the uncertainty in the distribution of births. This approximation will be most accurate when the number of births is large.
To summarize, under a plausible sampling model, the distribution of the observed SBH data is a mixture of a convolution of binomials, but with a number of approximations, we obtain the closed-form likelihood, Eq.~(\ref{eq:obslik}), that is far more straightforward to work with. Within this likelihood, there are two sets of parameters to estimate, those associated with birth ($ \bc$) and those with death ($\bq$).

From (\ref{eq:obslik}) we note that,
\begin{align}
E[T_D^{m_s} |  T_B^{m_s}, \bc, \bq] & = E[E\{T_D^{m_s} | \bB^{m_s}, \bc, \bq\}] \nonumber\\
& = E\left[ \sum_{a=0}^{m_s} B_aq(a)\big| T_B^{m_s},\bc, \bq \right]  \nonumber \\
& = T_B^{m_s} \sum_{a=0}^{m_s}c_{m_s}(a) q(a), \label{eq:brass}
\end{align}
which is the expectation of Eq.~(\ref{eq:obslik}).
The Brass method essentially treats Eq.~(\ref{eq:brass}) as deterministic, replacing the left side with the observed number of deaths. 

In this section we have shown that under a plausible and flexible model, the sampling model for the data available in SBH is a mixture of a convolution of binomial distributions, but by replacing the binomials by Poissons and approximating the distribution of births, we obtain a tractable sampling model. This form contains two sets of probabilities, $c(\cdot)$ for births and $q(\cdot)$ for deaths, and each of these may be modeled as functions of covariates and space and time, as we now describe.

%
%

\subsection{The Full Model}

%
The derivation in the last section will now be extended to the more realistic scenario in which the births and mortality models have greater complexity. We provide a summary here, with full details relegated to the Supplementary Materials. We specify models for fertility and mortality.
Let $f_m(\bx_t)$ denote the probability a woman gives birth at age $m$ and in year $t$ with $\bx_t$ containing the covariates at time $t$ associated with birth.
For mortality, we use a discrete hazards model. Let $_1 q_a(\bx_t) = q_a(1,\bx_t)$ denote the risk of mortality, i.e.,~the probability that a child dies between age $a$ and $a+1$ with $\bx_t$ now containing the covariates at time $t$ associated with mortality. Although we use $\bx_t$ for covariates in both models for notational convenience, the covariates used for each will generally differ. The parameter of interest is $q(5,\bx_t)$, the probability of death within 5 years of life, {at time} $t$ and with covariates $\bx_t$. This gives the probability of death before age 5 as,
\begin{align}
q(5,\bx_t) = 1 - \prod_{i=0}^{4}(1 -\, _1q_i(\bx_t)). \label{eq:trueq5}
\end{align}
 Let $q^\star(a, \bx_t)$ be the probability of dying within $a$ years {given birth at time} $t$ and with covariates $\bx_t$; thus,
 \begin{align}
 q^\star(a,\bx_t) = 1 - \prod_{i=0}^{a-1}(1 -\, _1q_i(\bx_{t+i})). \label{eq:starq5}
 \end{align}
The difference between Eqs.~(\ref{eq:trueq5}) and (\ref{eq:starq5}) for $a=5$ is subtle, but crucial. In Eq.~(\ref{eq:trueq5}), we envisage a synthetic cohort of children that are born in year $t$ and then repeat year $t$  five times at different ages, meaning they experience in the same calendar year the mortalities of each age band. In  Eq.~(\ref{eq:starq5}), we instead imagine a real cohort of children that are born in year $t$ and are followed up to year $t+5$.

For women who are of age $m_s$ years at the time of the survey, define $T_B^{m_s}(\bx_t)$ to be the total number of children ever born to  those women, with covariates $\bx_t$ at time $t$.   Similarly, define $T_D^{m_s}(\bx_t)$ to be the total number of children that ever died to women who are $m_s$ at the time of the survey with covariates $\bx_t$.

In the Supplementary Materials we derive a full likelihood, which is based on fertility and mortality models. 
Under a number of carefully defined simplifying approximations, we obtain a distribution for the observed deaths, $T_D^{m_s}$, given the observed births $T_B^{m_s}$ and dependent on birth parameters $\bc$ and mortality parameters $\bq^\star$:
\begin{align}
T_D^{m_s} |  T_B^{m_s}, \bc, \bq^\star& \sim  \text{Poisson}\left(T_B^{m_s}\sum_{a=0}^{m_s}c_{m_s}(a,\bx_{t_s-a}) q^\star(a)\right), \label{eq:sbhsimplified2}
\end{align}
where
$$
c_{m_s}(a,\bx_{t_s-a})  = \frac{f_{m_s -a}(\bx_{t_s-a})}{\sum_{a=0}^{m_s}f_{m_s -a}( \bx_{t_s-a})}. \nonumber
$$
Again, the total deaths is a mixture over the potential times at which the deaths could have occurred, with the mixture depending on the frequencies of births at different ages $c_{m_s}(a,\bx_{t_s-a})$ (given a birth occurring) and the probabilities of dying at different ages $q^\star(a)$. In practice, we specify models for the birth probabilities $\bc$ and the mortality parameters $\bq^\star$. Examples of such models are given in both the simulation study and the substantive application to estimating U5MR in Malawi.

Since $\bc$ is often unknown, we propose first fitting a fertility model to the FBH data. Such a model allows an estimate of $f(m,\bx_t)$ to be formed, i.e.,~$\widehat{f}(m,\bx_t)$, which can then be transformed to $\widehat{c}_{m_s}(a,\bx_t)$. In the simulation study that we next describe, we  investigate the implications of the approximations embedded in (\ref{eq:sbhsimplified2}).

\section{Simulation Study}

In the simulation, we suppose  there are two surveys that contain birth history information, both taken in 2010. One survey provides FBH information and a much larger survey (analogous to the census in our Malawi application) provides SBH information. We used the geography of Kenya, which is comprised of $47$ counties, labeled by $r=1,\dots, 47$. In total, we simulated FBH data for $47 \times 4,000$ women and SBH data for $47 \times 20,000$ women, with equal numbers in each region. 
The Supplementary Materials give details on how the data was simulated and also the birth probabilities, $f_m$,  used. 
These probabilities were set to be constant over 5-year age groups of women and closely resemble patterns observed in the 2010 Malawi DHS 
\citep{MalawiDHS:10}.
The birth probabilities  were also assumed constant over space and time. 
That is,
$$
f_m(\bx_t)  = \frac{\exp(\beta_{c[m]})}{1+\exp(\beta_{c[m]})},
$$
with $c(m)$ an indicator function for mother's age $m$ that takes on values: 1 if $m=15,\dots,19$, 2 if $m=20,\dots,24$, 3 if $ m=25,\dots,29$, 4 if  $m=30,\dots,34$ and 5 if $ m=35,\dots,49$. Hence, we model the birth probabilities as being a five-level factor variable, with $\exp(\beta_{c})$ being the odds of a birth in age band $c$, for $c=1,\dots,5$.
We simulate deaths using three distinct discrete hazards: one for the first year of life, one for years 1,2,3,4, and one for 5 onwards. Each of these are a function of time (5-year periods) $t$ and region $r$, so that,
\begin{align*}
_1q_a(t,r)& = \frac{\exp(\beta_{c[a]} + \phi_{c[a]}(t) + S_r + \epsilon_r)}{1+\exp(\beta_{c[a]} + \phi_{c[a]}(t) + S_r + \epsilon_r)},
\end{align*}
where $c[a]$ takes the values: 1 for $\{a=0\}$, 2 for $\{a=1,\dots,4\}$, 3 for $\{a=5,\dots\}$. The interpretation of the parameters is as follows:
 \begin{itemize}
 \item $\exp(\beta_{c})$ are the odds of death at time 0, and are taken as fixed effects, with $c=1,2,3$, to allow the age curves to start from 3 different intercepts. 
 \item $\exp(\phi_{c}(t))$ describe how the odds of death in age band $c$ change across years, via a smoothing model. Specifically, the three $\phi_c(t)$ terms ($c=1,2,3$) are modeled as second order random walks (RW2) over time  \citep{rue:knorrheld:05}.  In terms of second differences:
\begin{equation}\label{eq:rw2}
\underbrace{\left[~ \phi_{c}(t) -\phi_{c}(t-1) ~\right] }_{\text{Slope between $t$ and $t-1$}}- 
\underbrace{\left[~ \phi_{c}(t-1)-\phi_{c}(t-2) ~\right] }_{\text{Slope between $t-1$ and $t-2$}}
\sim N(~0, 1/\kappa_T~),
\end{equation}
showing that deviations from linearity are being modeled, i.e.,~this model encourages a linear trend for the log odds, but allows fluctuations if the data suggests these are warranted. The model has a single smoothing parameter, the precision $\kappa_T$, with large values giving smoother trajectories. In (\ref{eq:rw2}), if $\kappa_T$ is large, the variance is small and so the local slopes are more tightly tied together, meaning they are similar in magnitude. We emphasize that the parameter $\kappa_T$ is assigned a prior and then estimated from the data, so that an appropriate amount of smoothing is applied.

 \item $\exp(S_r)$ describe how the odds of death vary smoothly across geographical regions $r$. 
 The spatial smoothness is modeled via an intrinsic conditional autoregressive (ICAR) model \citep{rue:knorrheld:05}. Under this model, the spatial effects $S_r$ are modeled { conditional} on the neighbors. Specifically,
$$S_r | \{ S_{r'} = s_r, r' \sim r \}, \kappa_S  \sim N\left( \overline{s}_r, \frac{1}{\kappa_S m_r}\right),$$
where $r' \sim r$ is shorthand for $r'$ is a neighbor of $r$, and where we define a neighboring area  as one that shares a boundary, $\overline{s}_r =\frac{1}{m_r} \sum_{r' \sim r} s_r$ is the mean of the neighbors  of area $r$ and $m_r$ is the number of such neighbors. The effect of this prior is to pull mortality risks toward neighboring risks, so that $\kappa_S$ is a spatial smoothing parameter with large values indicating little spatial smoothing.
\item $\exp(\epsilon_r)$ is an independent odds contribution at the region level that corresponds to a ``random shock", in that there is no spatial structure. The random regional shocks are modeled as $\epsilon_r \sim_{iid} N(0, 1/\kappa_\epsilon)$ with iid shorthand for ``independent and identically distributed". Hence, if $\kappa_\epsilon$ is large there are small random shocks only. 
 \end{itemize}
 Together, $S_r$ and $\epsilon_r$ correspond to the celebrated Besag, York, Molli{\'e} (BYM) model, that was introduced in \cite{besag:etal:91}, and is the most commonly used model in spatial epidemiological studies. To summarize, both the RW2 and ICAR terms encourage local smoothness in the mortality risk in time and space, respectively. Each have a sum-to-zero constraint for identifiability. For the simulation we used values of $\phi_{c}(t)$ that were similar to those observed in our Malawi application, and simulated the structured ($S_r$) and unstructured ($\epsilon_r$) random effects to produce spatial risk surfaces. 

We examined model-fitting when three different types of birth information are available  (see Supplementary Materials for more details):
\begin{enumerate}
\item The true number of SBH births by age are known, $B_{m_s}(a)$,  and included in the Poisson model. 
\item The true birth probabilities, $c_{m_s}(a)$, are known and the approximate Poisson model (Eq.~(\ref{eq:sbhsimplified2})) is used to include the SBH data.
\item The birth probabilities are estimated from the FBH data  via logistic regression, and the estimates $\hat{c}_{m_s}(a)$ are plugged in for $c_{m_s}(a)$, in the approximate Poisson model (Eq.~(\ref{eq:sbhsimplified2})).
\end{enumerate}
The Supplementary Materials include a plot of the observed fertility in the SBH data, along with the various approximations. At all ages we see that the results are almost identical under the different model-fitting strategies.

\begin{figure}[tbp]
    \centering
    \includegraphics[width=0.8\linewidth]{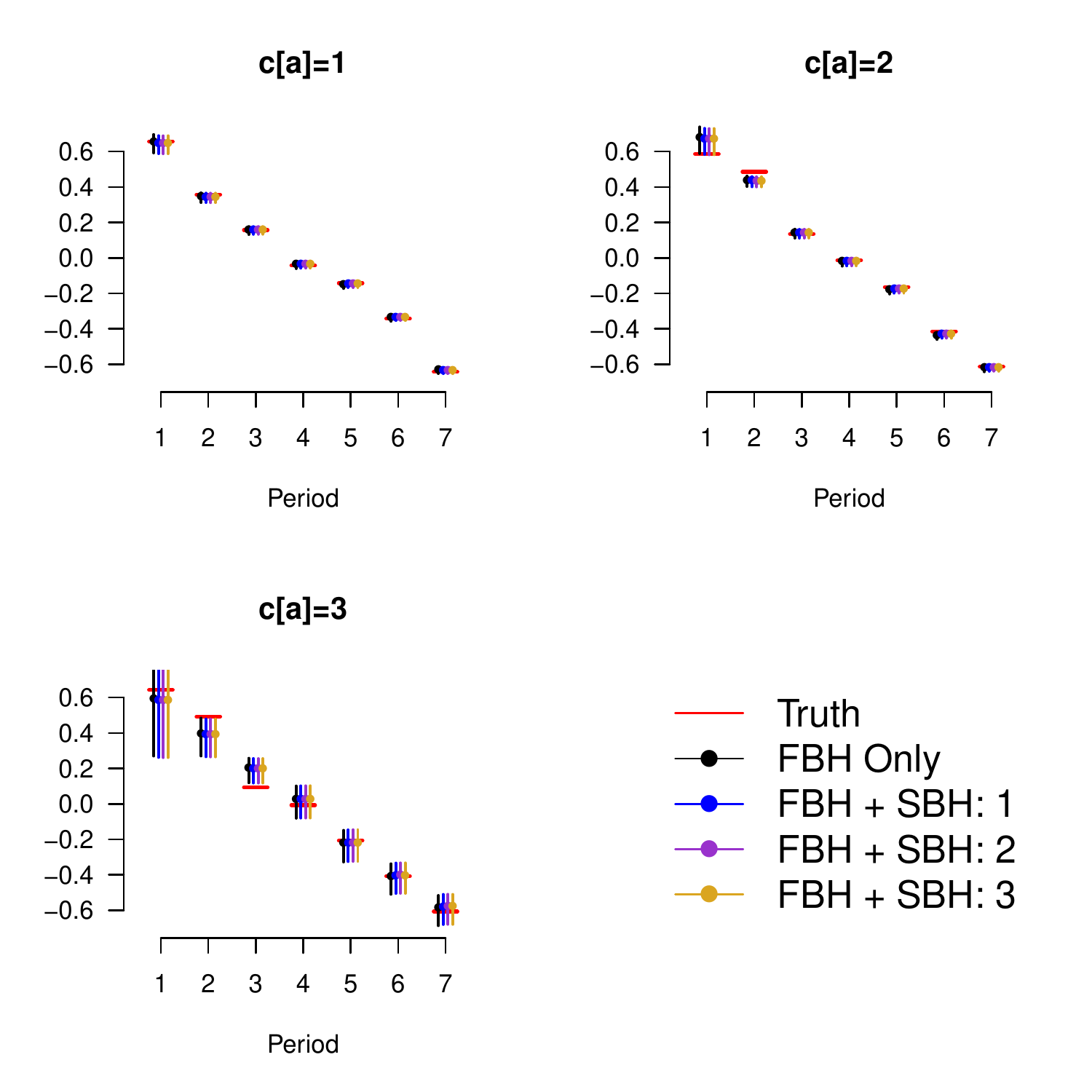}
    \caption[Mortality RW2 parameters in simulation]{Simulation study summaries: Posterior median estimates of RW2 parameters, $\phi_{c[a]}(t)$, in mortality model, with 95\% uncertainty intervals, for three different age groups $c[a]$ (0--1, 2--5, 5+). We include the truth, FBH only and three approaches for dealing with the SBH data: (1) true births, (2) true birth probabilities, (3) estimated birth probabilities.}
    \label{fig:bv:rw2sim}
\end{figure}

Table  \ref{tab:mort-sim} reports the true parameter values and the estimates and uncertainty for the intercepts and smoothing parameters under a FBH only analysis, and under combined FBH+SBH analyses in which different information was used on births/fertilities. Figure \ref{fig:bv:rw2sim} displays how the random walk temporal smoothing estimates for mortality, $\widehat \phi_{c}(t)$,  evolve over time $t$ under the different approaches.
The three approaches to incorporating SBH yield very similar results and are comparable to the FBH only summaries. All models underestimate the true intercept for the youngest age group 0--1. This is expected as the Poisson approximation to the Binomial is best when the probability of the event is small and mortality is highest in the first year of life. Also, there exists some identifiability problems with the two spatial random effects $S_r$ and $\epsilon_r$ (which is a common phenomena in spatial analyses, and not a problem when prediction of the overall mean, rather than the separate components is all that is required); hence, we display results for $S_r + \epsilon_r$ in the Supplementary Materials.

\begin{table}[!h]
    \footnotesize
    \centering
    \begin{tabular}{c|lllll}
       & True & FBH Only & FBH + SBH: 1 & FBH + SBH: 2 & FBH + SBH: 3\\
     \hline
    $\exp(\beta_0)$ & 0.150 & 0.148 (0.144, 0.153) & 0.148 (0.143, 0.153) & 0.148 (0.143, 0.153) & 0.148 (0.143, 0.153)\\
    $\exp(\beta_1)$ & 0.053 & 0.054 (0.052, 0.055) & 0.053 (0.051, 0.055) & 0.053 (0.051, 0.055) & 0.053 (0.051, 0.055)\\
    $\exp(\beta_2)$ & 0.005  & 0.005 (0.005, 0.006) & 0.005 (0.005, 0.006) & 0.005 (0.005, 0.006) & 0.005 (0.005, 0.006)\\
    $\kappa_T$ & --  & 200 (73, 542) & 199 (75, 531) & 197 (73, 539) & 198 (74, 539)\\
    $\kappa_S$ & 45  & 34 (3.5, 313) & 38 (4.0, 342) & 37 (3.9, 350) & 37 (3.7, 365)\\
    $\kappa_{\epsilon}$ & 90 & 103 (3.85, 2760) & 86 (5.11, 1540) & 86 (4.8, 1540) & 87 (5.1, 1510)
    \end{tabular}
    \caption{Simulation study summaries: Comparison of estimates and 95\% uncertainty intervals  when using FBH only and FBH + SBH data where SBH data is incorporated using one of the three approaches: (1) true births, (2) true fertilities, (3) estimated fertilities. The exponentiated intercepts (odds ratios) are $\exp(\beta_0), \exp(\beta_1), \exp(\beta_2)$ and the precision (smoothing parameters) for time, space and iid terms are $\kappa_T$, $\kappa_S$, $\kappa_\epsilon$. The time trend was taken to mimic the Malawi data (so that we do not simulate from a RW2 model), and so there is no true precision parameter on the random walk.}
    \label{tab:mort-sim}
\end{table}

\begin{figure}[tbp]
     \centering
     \includegraphics[width=0.85\linewidth]{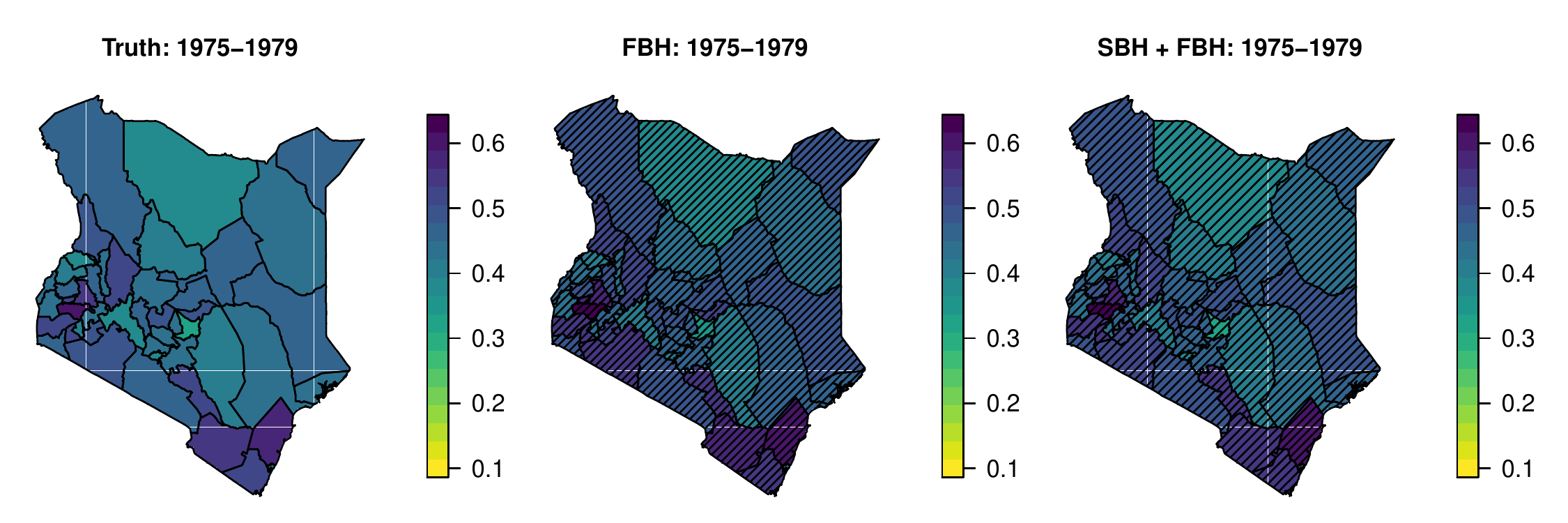}\\
     \includegraphics[width=0.85\linewidth]{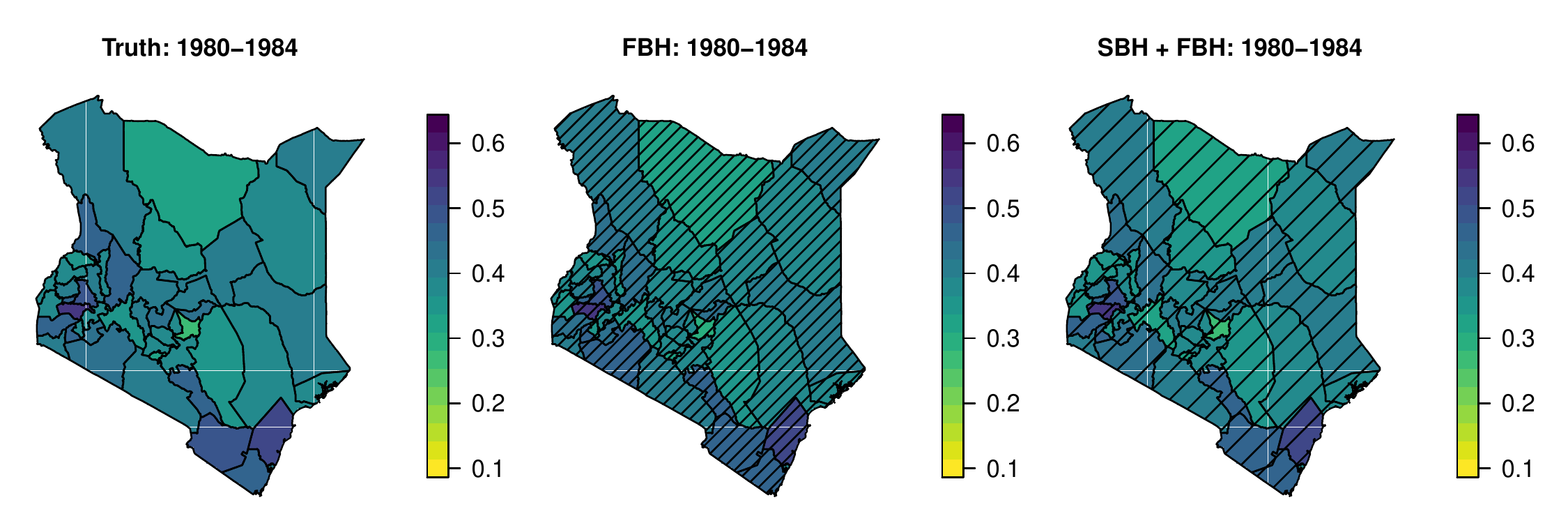}\\
     \includegraphics[width=0.85\linewidth]{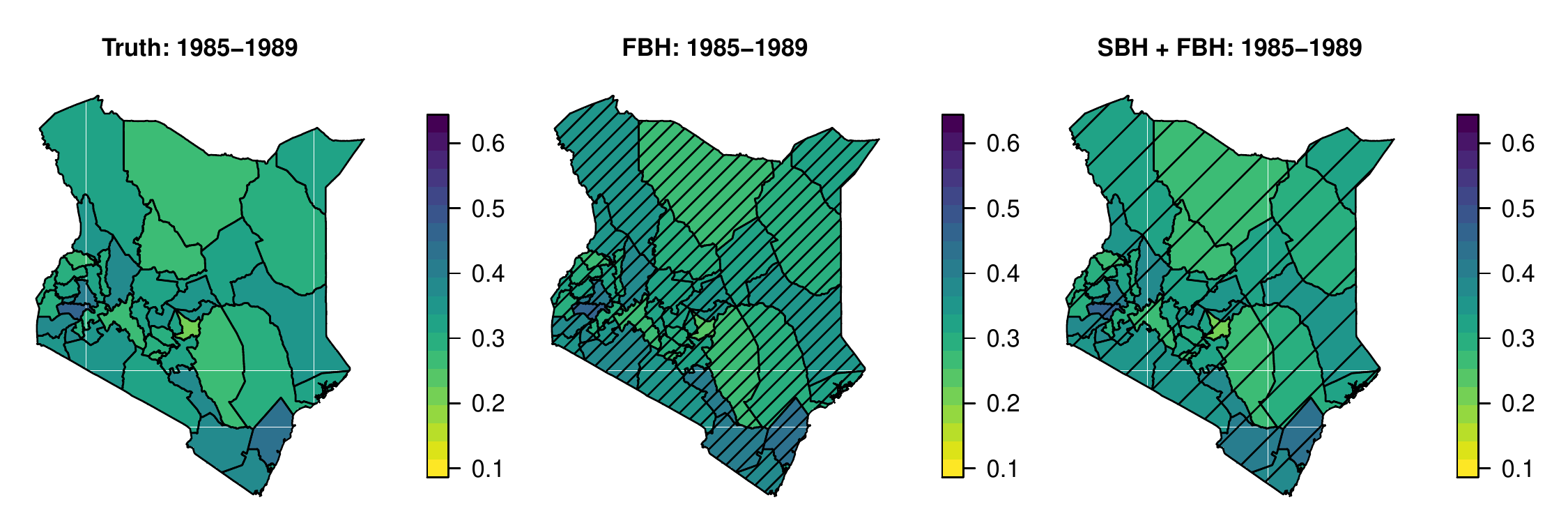}\\
     \includegraphics[width=0.85\linewidth]{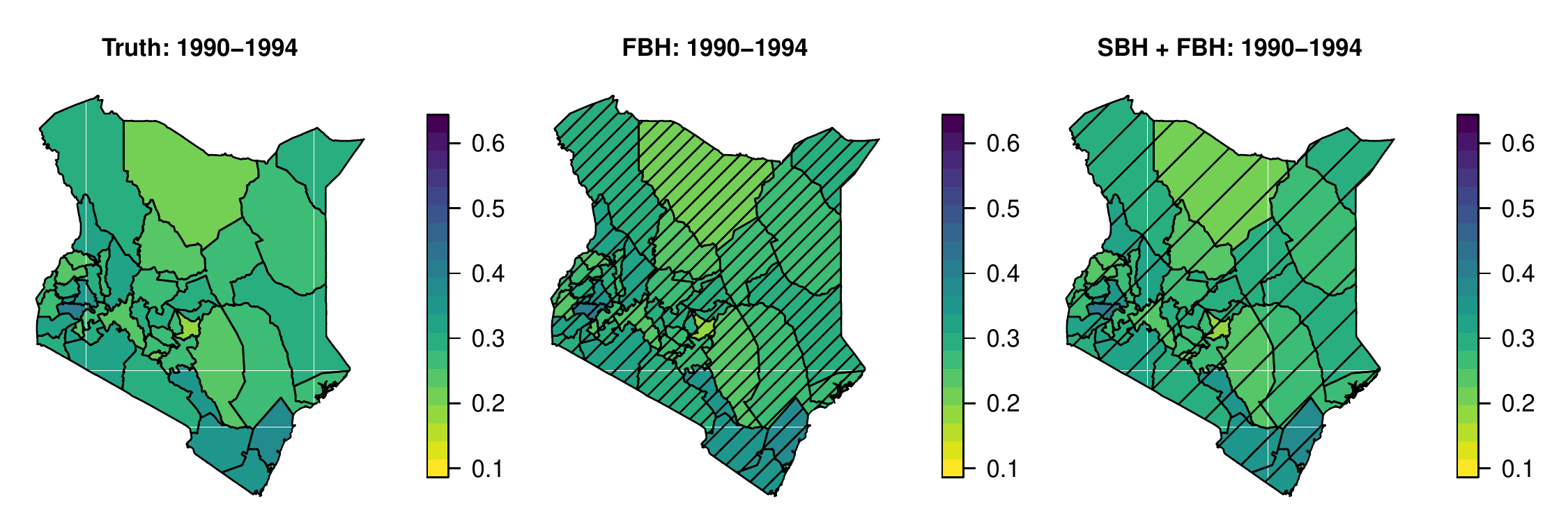}
    \caption[True and estimates of U5MR in the first four periods in Kenya simulation]{Simulation study summaries: True and estimates of U5MR in the first four periods by Kenyan counties. Uncertainty (standard deviation of logit U5MR) is expressed through hatching with denser hatching indicating greater uncertainty. Note the wider hatching in the right hand column, as compared to the middle column, which shows the added benefit of the SBH data. The hatching legend is given in Figure \ref{fig:bv:q5bsim}.}
    \label{fig:bv:q5asim}
\end{figure}

\begin{figure}[tbp]
    \centering
     \includegraphics[width=0.85\linewidth]{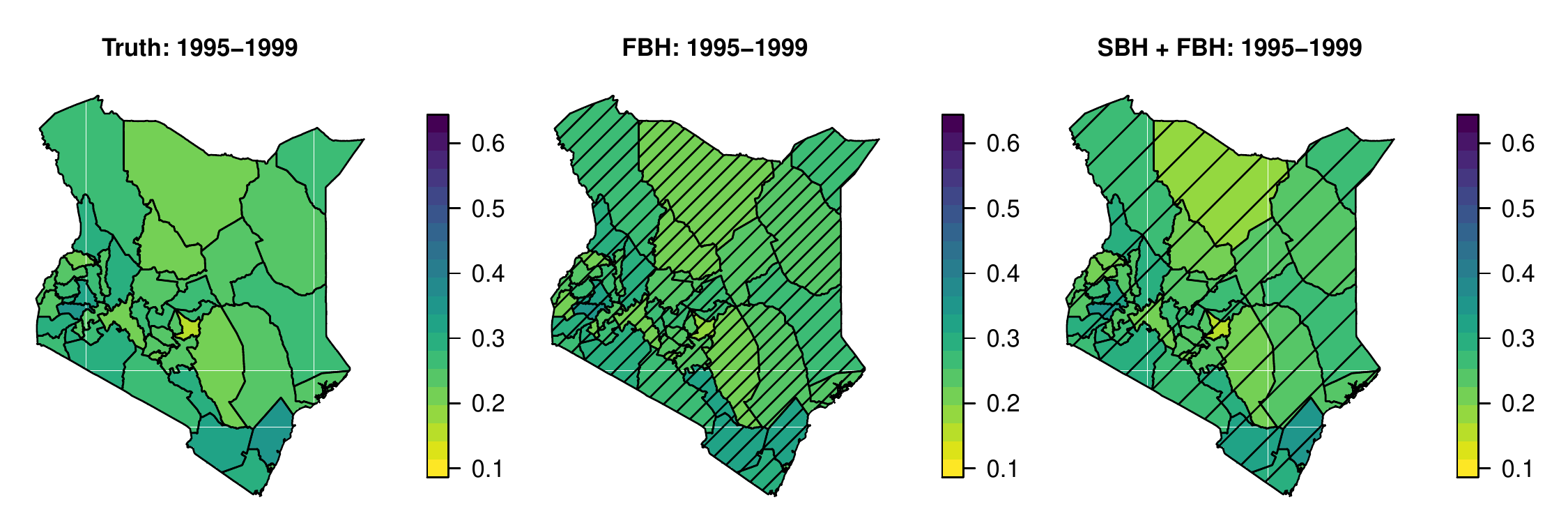}\\
     \includegraphics[width=0.85\linewidth]{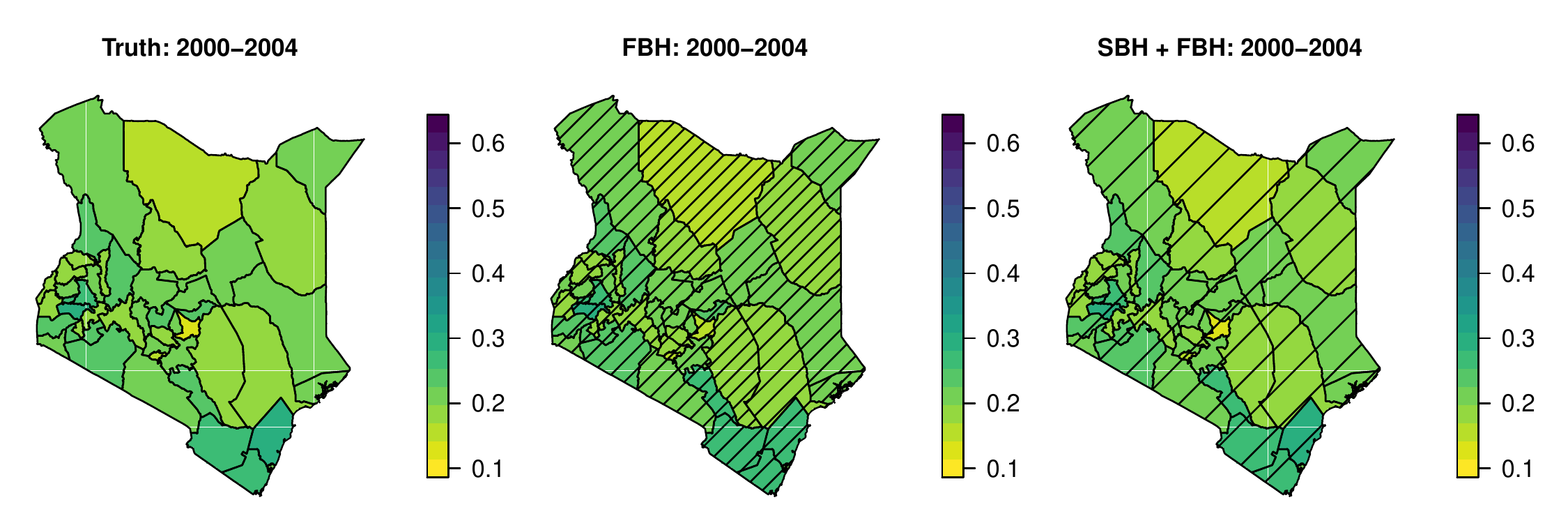}\\
     \includegraphics[width=0.85\linewidth]{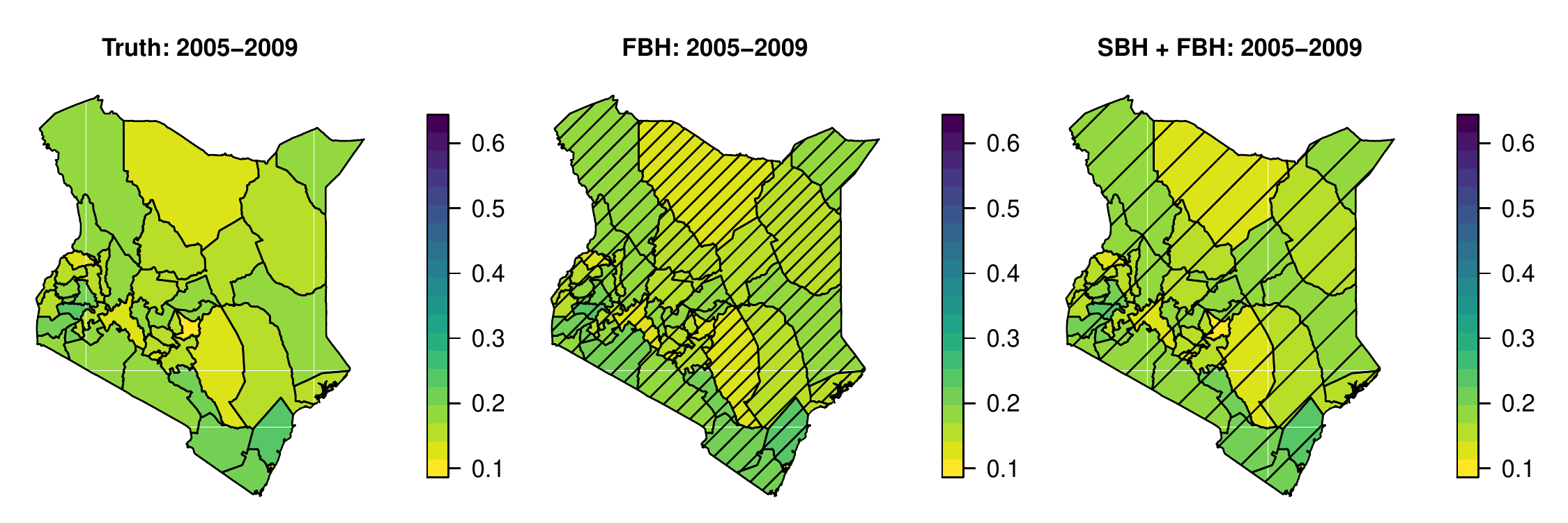}\\
     \includegraphics[width=0.45\linewidth]{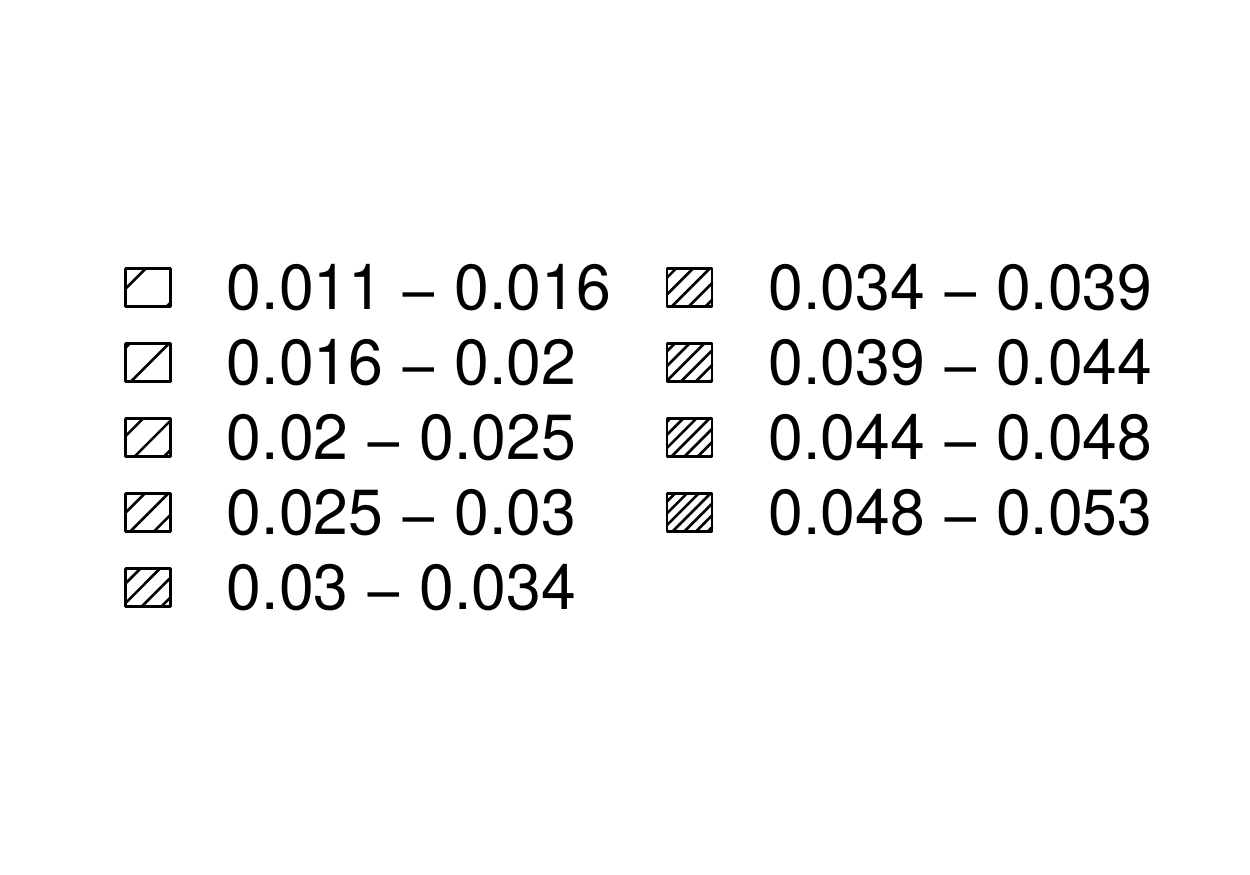}
    \caption[True and estimates of U5MR in the last three periods in Kenya simulation]{Simulation study summaries: True and estimates of the U5MR in the three most recent periods by Kenyan counties. Uncertainty (standard deviation of logit U5MR) is expressed through hatching with denser hatching indicating greater uncertainty. Note the wider hatching in the right hand column, as compared to the middle column, which shows the added benefit of the SBH data.}
    \label{fig:bv:q5bsim}
\end{figure}

The mortality hazard can be combined to give U5MR estimates over time via Eq.~(\ref{eq:trueq5}).
Figures \ref{fig:bv:q5asim} and \ref{fig:bv:q5bsim} visually depict estimates of U5MR with uncertainty (conveyed via the standard deviation), expressed using hatching; denser hatching reflecting greater uncertainty. See the Supplementary Materials for more details on how these estimates were obtained. For presentation purposes, ``SBH + FBH'' refers to using the third approach for including the SBH data (i.e.,~plugging in the fertility estimates from the FBH only analysis). Both the FBH and SBH+FBH give similar estimates for U5MR and are close to the truth, but we clearly see that the uncertainty is reduced when SBH data is incorporated, as expected. For reference, the estimated standard deviation (on the logit U5MR scale) in 1975--1979 was 10\%--21\% (mean 14\%) higher than when only FBH data was used. In 2005--2009, the estimated standard deviation was 90\%--120\% (mean 105\%) higher when only FBH data was used, as compared to the analysis with FBH+SBH data. 

We conclude from this simulation that the approximations perform well in general, and the SBH data can produce more precise estimates when used to supplement FBH data, when compared to estimates from the FBH data alone.

\section{Application to Malawi Data}

\subsection{Context}

In the 5-year period before the 1992 Malawi Demographic and Health Survey (DHS) the national U5MR was 234 deaths per 1,000 live births, and this fell to 63 deaths per 1,000 in the 5-year period prior to the 2015-16 DHS \citep{malawi2016}. This impressive fall hides significant subnational variation, however. Using simple weighted estimation techniques the most recent DHS and Multiple Indicator Cluster Surveys (MICS) reported summaries for three sub-regions of Malawi. Table CM.2  of \cite{malawimics2014} gives U5MR estimates for Northern, Central and Southern regions (over the 5 years prior  to the survey) as 67, 81 and 92, respectively. Similarly, in Table 8.2 of \cite{malawi2016} the U5MR estimates for Northern, Central and Southern regions (over the 10 years prior to the survey) are 57, 81, and 73. Even between these two surveys we see inconsistencies of estimates in the Southern region.

 In this analysis we estimate 5-yearly U5MR in each of the districts of Malawi, using FBH and SBH data. We note that although there are 28 districts, due to the sampling done in the 2006 MICS we combine some of the districts so that we end up modeling 26 regions. Specifically, Mwanza and Neno were combined into a single region as were Nkhata Bay and Likoma.

\subsection{Data}

The available data includes three DHS, taken in 2004, 2010, and 2015; two MICS, taken in 2006 and 2013; and a census taken in 2008. All five surveys contain FBH data, whereas the census contains SBH data only.  Microdata, from which we can access the required geographic information, is available on a 10\% random sample of the census.  Table \ref{tab:malawidat} summarizes the numbers of clusters, women, births and deaths for the Malawi data. We see the census data dominates in terms of women, births and deaths, but the lack of times on births and deaths, means the information content of these data is far less than that provided by the FBH data. The cluster surveys are stratified by geographical area and urban/rural. The latter means it is important to include a term for urban/rural in the model, to account for the differential sampling of urban and rural clusters (for example, in the 2015 Malawi DHS, urban clusters were almost twice as likely to be sampled as rural clusters), and the differences in mortality between urban and rural areas. However, urban residence is defined at the time of survey, not at the time of exposure to risk of births or deaths, and so for periods well before a survey any urban effect is likely to be diluted by migration and urbanization.

\begin{table}[!h]
\centering
\begin{tabular}{l|rrrr}
Survey &\# Clusters & \# Women & \# Births & \# Deaths \\
\hline
DHS 2004 & 521 & 11,698 & 35,883 & 6,534 \\
MICS 2006 & 1,040 & 26,211 & 78,641 & 11,855 \\
Census 2008 &-- & 309,851 & 875,423 & 150,798 \\
DHS 2010 & 849 & 23,020 & 72,301 & 11,343 \\
MICS 2013 & 1,139 & 24,220 & 72,568 & 9,213 \\
DHS 2015 & 850 & 24,562 & 68,074 & 7,235
\end{tabular}
\caption{Data summaries for Malawi. SBH data is available from the census, and FBH from all other data sources (which are all household surveys).}\label{tab:malawidat}
\end{table}

In the Supplementary Materials we present results from an exploratory analysis, comparing the five surveys and census. These results suggest that women in the census tend to report similar total number of children born as in the five surveys; however, women in rural areas tended to report that more of their children died. This observation motivated inclusion of a bias term by urban/rural in our mortality model.

The fertility model we use is
\begin{align*}
f(m, \bx_t) = \frac{\exp(
\beta_m + \beta_{\text{URB}} \times 1(\mbox{urban})+ \phi_{c[m]}(p) + S_r+ \epsilon_r)}
{1+\exp(
\beta_m + \beta_{\text{URB}} \times 1(\mbox{urban})+ \phi_{c[m]}(p) + S_r+ \epsilon_r)}
\end{align*}
where $\beta_{\text{URB}}$ is a fixed effect for strata (urban/rural), with $1(\mbox{urban})$ being an indicator for whether the mother resides in an area classified as urban, $\beta_m$ are fixed effects for mother's age (for mothers in areas classified as rural),  $\phi_{c[m]}(p)$ is a mother's age group specific RW2 in roughly 5-year time periods $p$, $S_r$ is an ICAR spatial random effect and $\epsilon_r$ is an unstructured (iid)  error on regions $r$ that allows for independent ``shocks".

The mortality model is,
\begin{small}
\begin{align*}
 _1q_a(\bx_t) & = \frac{\text{HIV}(p)\exp( \beta_{\text{URB}}  \times 1(\mbox{urban}) + \beta_{\text{SBH}}  \times 1(\mbox{SBH})  + \beta_{\text{SBH,URB}} \times 1(\mbox{SBH and urban}) 
+ \beta_{b[a]} + \mbox{smooth})}{1+\text{HIV}(p)\exp( \beta_{\text{URB}}  \times 1(\mbox{urban}) + \beta_{\text{SBH}}  \times 1(\mbox{SBH})  + \beta_{\text{SBH,URB}} \times 1(\mbox{SBH and urban}) 
+ \beta_{b[a]} + \mbox{smooth})}
\end{align*}
\end{small}
where
$$\mbox{smooth} =  \phi_{c[a]}(p) +  S_r + \epsilon_{rt}.
$$
This model therefore has mortality being different in urban/rural areas and allows for systematic bias in the SBH data, as compared to the FBH data (where we are assuming that the FBH are less likely to be biased than the SBH data), and the size of the bias can be different in urban and rural areas. These associations are assumed to be constant across region and time period so that the overall levels are not informed by the SBH data, but rather the data informs on the spatial and temporal differences. 


Age-specific intercepts and age-specific random walks, are indexed by,
\begin{align*}
b[a] = \begin{cases}
0 & a=0,\\
1 & a=1,\\
2 & a=2,\\
3 & a=3,\\
4 & a=4\\
5 & a=5,6,\dots
\end{cases} \qquad
c[a]  = \begin{cases}
0 & a=0,\\
1 & a=1,\dots,4\\
2 & a=5,6,\dots
\end{cases}
\end{align*}
respectively. So the curves start from six different points (defined by different ages), but subsequently follow three curves over time; ages 1,2,3,4 share a trajectory, for reasons of parsimony, and based on initial analyses. Hence, these four age trajectories are parallel.
We again include ICAR spatial random effects $S_r$ for region along with iid errors $\epsilon_{rt}$ crossed over region and time. Although we use the same letters for these spatial terms in both models, we stress that we have separate spatial and iid terms for each of fertility and mortality.
The fixed effects $\beta_{b}$ are yearly intercepts for the first 5 years, with an additional term for all ages greater than 5. if there were interest in yearly mortality beyond 5, we would include yearly intercepts for this range also (though deaths become much rarer after age 5). The data for Malawi is relatively extensive and so we do not smooth over age. In other situations in which there were less data it would be natural, and straightforward, to place either a smoothing prior, for example a RW2, on the $\beta_{b}$ paramegers, or to place  informative priors on these coefficients.
We include an HIV adjustment, as in previous work with FBH data \citep{wakefield:etal:19}. Briefly, we wish to adjust for the loss of child deaths due to mothers who have died of AIDS; the children of these mothers are more likely to die, and so the missingness is informative. We estimate the bias using the method described in \cite{walker:etal:12}, which uses a cohort component projection model.

\subsection{Results}

In the Supplementary Materials we give parameter summaries and visual summaries for fertility. For mortality, Table \ref{tab:mmort} gives the parameter estimates. As expected, mortality decreases monotonically with age, as indicated by the odds ratio estimates, $\exp(\beta_0)$--$\exp(\beta_5$). Mortality is decreased in urban areas with the estimate from the FBH data alone indicating a 29\% decrease.  
The SBH data gives 18\% higher mortality estimates in rural areas
 and approximately the same mortality estimates in urban areas as found in the SBH data. This finding was confirmed by examination of the raw data (see Supplementary Materials).

\begin{table}[!h]
    \footnotesize
    \centering
%
\begin{tabular}{c|ll}
       & FBH Only & FBH + SBH\\
     \hline
    $\exp(\beta_0)$ & 0.103 (0.097, 0.109) & 0.102 (0.095, 0.109)  \\
    $\exp(\beta_1)$ & 0.033 (0.031, 0.036) & 0.032 (0.029, 0.035)\\
    $\exp(\beta_2)$ & 0.025 (0.023, 0.027) & 0.024 (0.022, 0.026)\\
    $\exp(\beta_3)$ & 0.018 (0.016, 0.019) & 0.017 (0.015, 0.018)\\
    $\exp(\beta_4)$ & 0.009 (0.008, 0.010) & 0.009 (0.008, 0.010) \\
    $\exp(\beta_5)$ & 0.005 (0.004, 0.006) & 0.005 (0.004, 0.007) \\ \hline
    $\exp(\beta_{\text{URB}})$ & 0.710 (0.686, 0.736) & 0.718 (0.694, 0.743)\\
    $\exp(\beta_{\text{SBH}})$ &--  & 1.183 (1.169, 1.198)\\
    $\exp(\beta_{\text{SBH,URB}})$ & -- & 0.850 (0.817, 0.884)\\ \hline
    $\kappa_T$ & 24.9 (11.0, 57.1) & 9.04 (4.00, 20.2)\\
    $\kappa_S$ & 123 (52.1, 291) & 153 (66.5, 349)\\
    $\kappa_{\epsilon}$ & 322 (118, 888) & 417 (141, 1220)
    \end{tabular} 
    \caption{Malawi application. Comparison of parameter estimates and 95\% uncertainty intervals when using FBH only and FBH + SBH data. Exponentiated intercepts for the age trends are  $\exp(\beta_0)$--$\exp(\beta_ 5)$. Exponentiated intercept for urbanicity is $\exp(\beta_{\text{URB}})$. Exponentiated bias terms for the SBH data by urban/rural are $\exp(\beta_{\text{SBH}})$ and $\exp(\beta_{\text{SBH,URB}})$. Precision for the time, space, and iid terms are $\kappa_T$, $\kappa_S$, and $\kappa_\epsilon$.}
    \label{tab:mmort}
\end{table}

The extent of the spatial and temporal smoothing can be determined by examination of the relevant smoothing parameters, with small values of the precisions (inverse variances) corresponding to stronger trends.
Here, we see that the time trends are very strong (low precision for RW2 model, $\kappa_T$).

\begin{figure}[tbp]
\centering
\includegraphics[width=\linewidth]{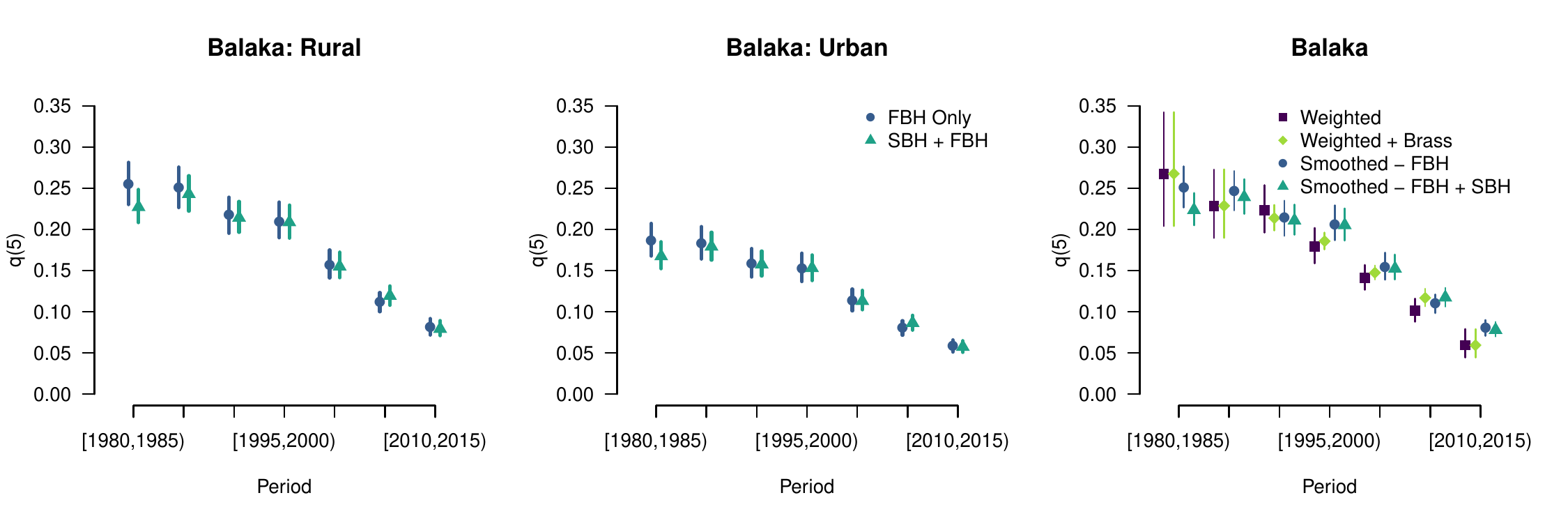}\\
\includegraphics[width=\linewidth]{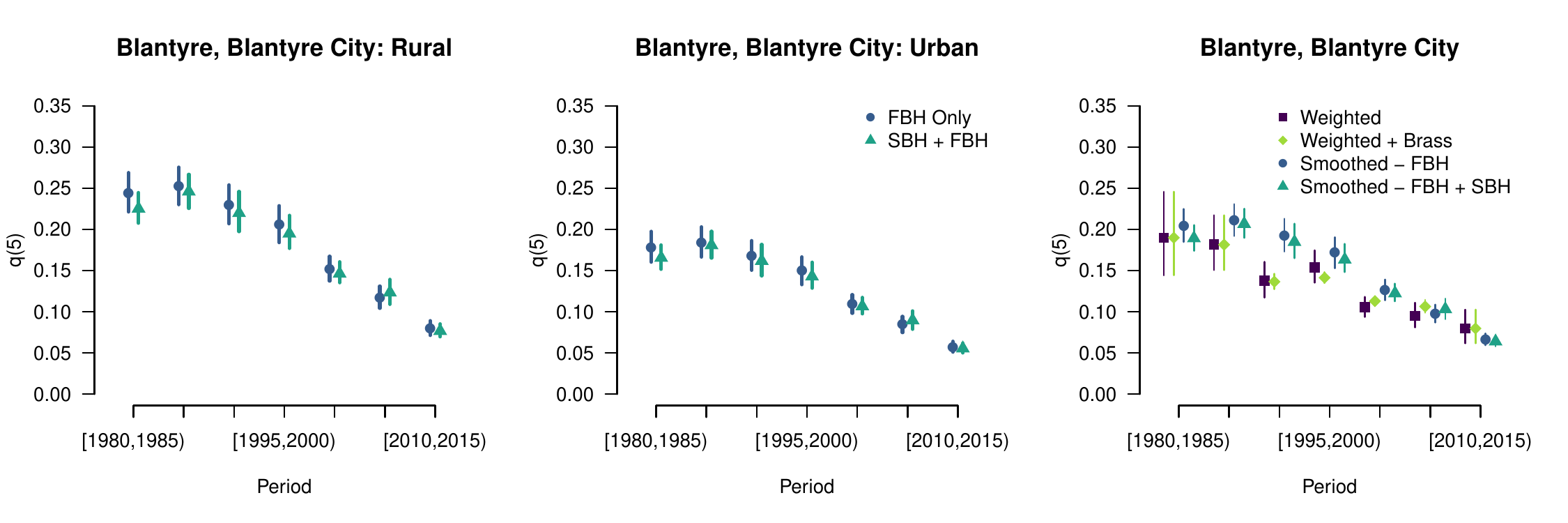}\\
\includegraphics[width=\linewidth]{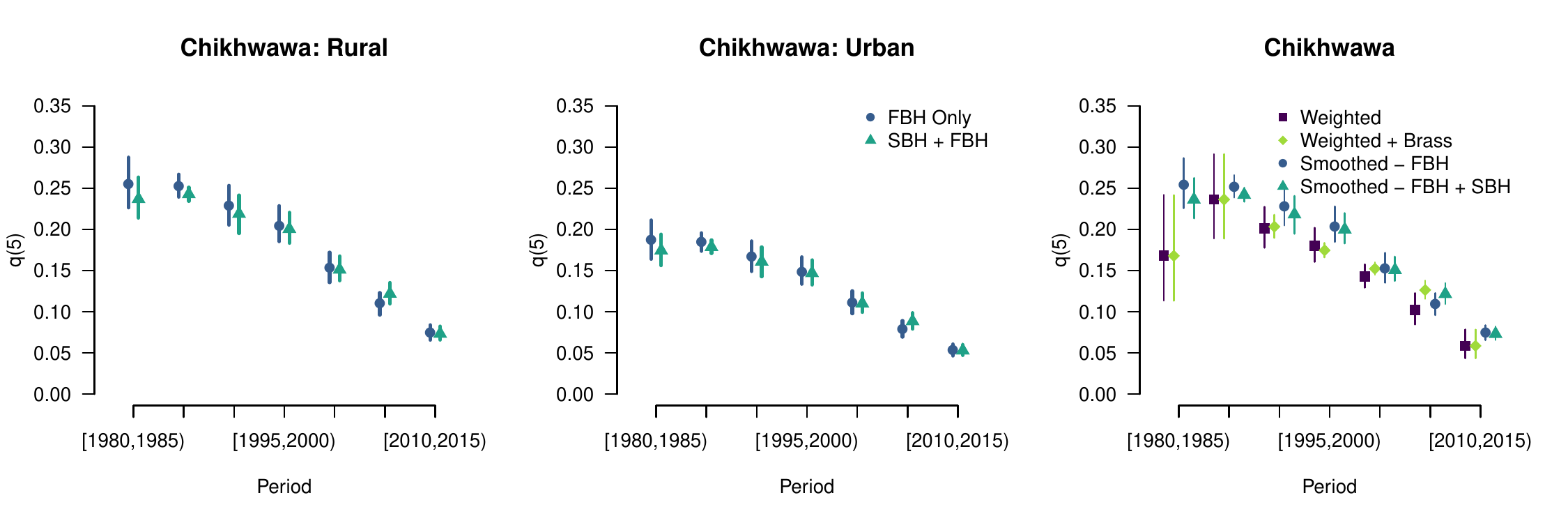}
\caption{Posterior medians and 95\% uncertainty intervals for U5MR in Malawi, stratified by urban and rural. Plots on the right are combined over strata and compare results with direct estimates and direct estimates combined with the Brass method.}
\label{fig:q5a}
\end{figure}

Figure \ref{fig:q5a} compares U5MR results, for three regions, using our model when only fit to FBH data and when we use our new approach to add in the SBH data. Results for the other 23 regions can be found in the Supplementary Materials. We also aggregate the results that are stratified by urban/rural up to the region level and compare to the direct estimates and the direct estimates combined with Brass.  In general, results are similar for all methods. Overall, U5MR decreases over time. Comparing FBH only and FBH + SBH results, U5MR tends to be lower in earlier time periods and higher in more recent time periods when SBH is included. Adding in SBH data with the Brass method also tends to result in higher U5MR in more recent time periods (as compared to results from only the direct estimates). The effect of the HIV pandemic can be seen in the slowing down of the decline in U5MR in the 1990s. The temporal trends are consistent  with national trends shown in Figure CM.3 of \cite{malawimics2014}.
 
Maps of the posterior median of U5MR are shown over time in Figure \ref{fig:q5mapa} (maps for the other time periods are in the Supplementary Materials). The density of the hatching corresponds to the standard deviation in (logit) U5MR, with denser hatching reflecting greater uncertainty. In general, we observe similar trends using FBH data alone and FBH + SBH, and a decrease in uncertainty overtime. Overall, uncertainty is reduced when the SBH data is included.

\begin{figure}[tbp]
\centering
\includegraphics[width=0.9\linewidth]{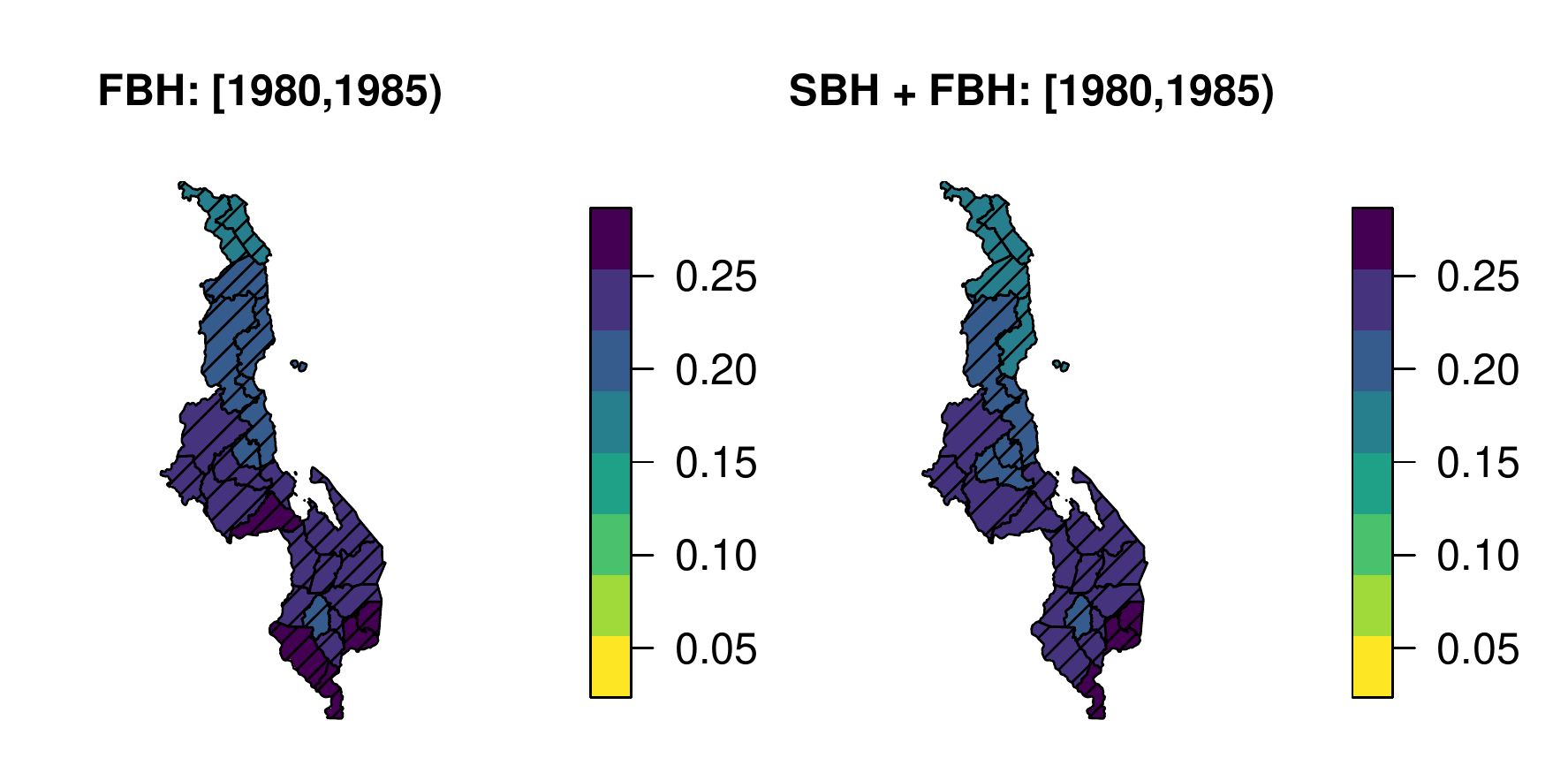}\\
\includegraphics[width=0.9\linewidth]{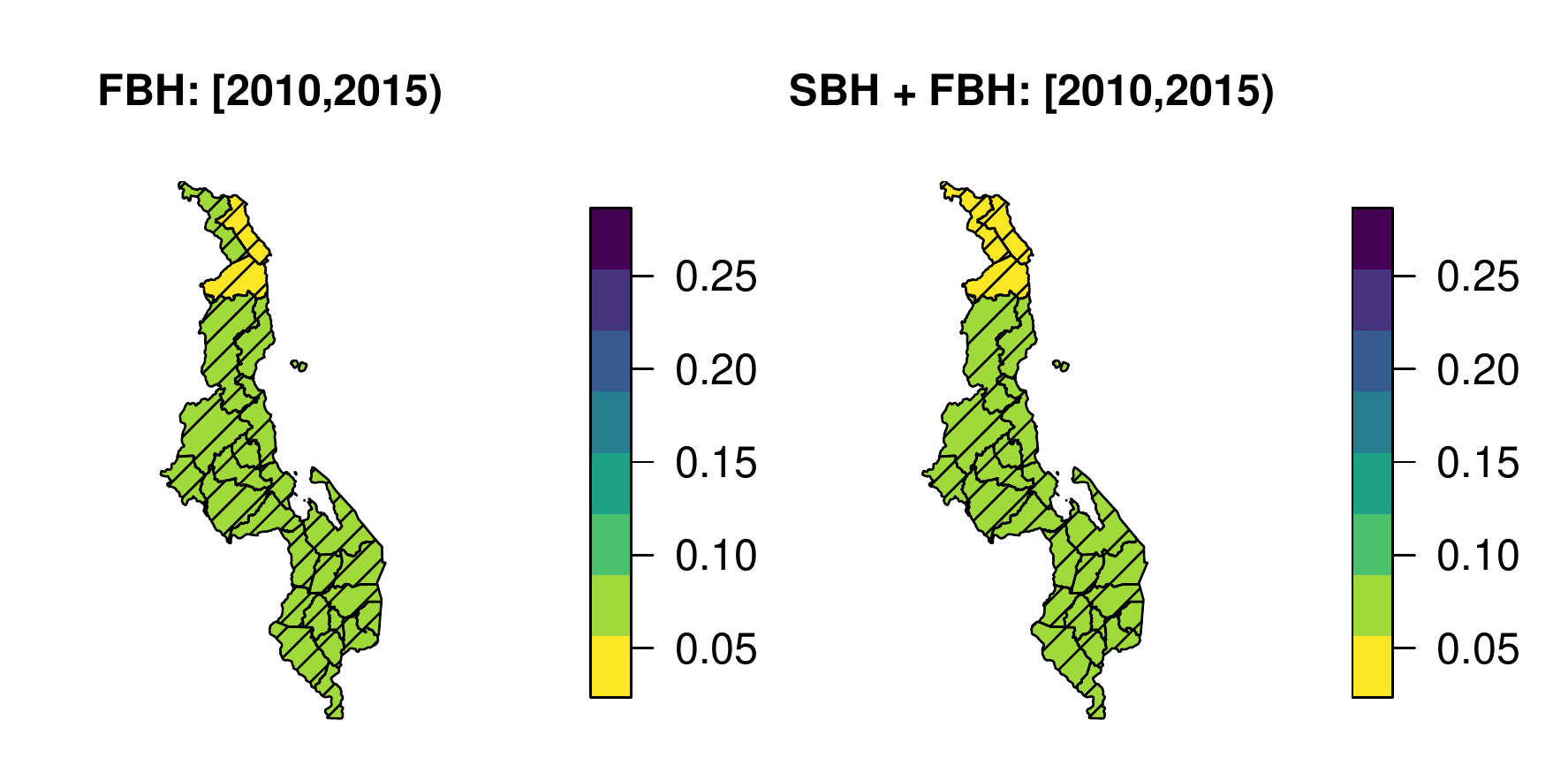}\\
\includegraphics[width=0.5\linewidth]{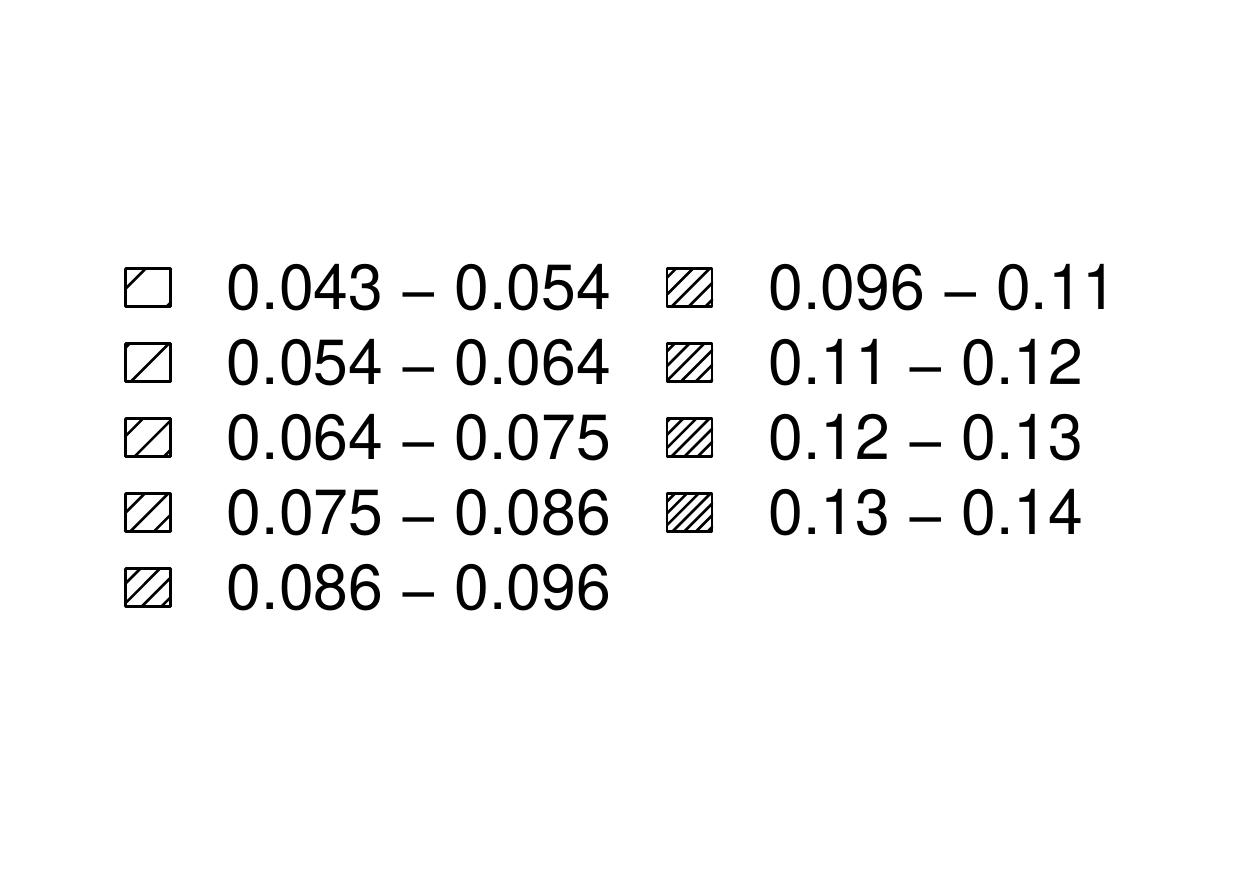}
\caption{Maps of U5MR (posterior medians) when using FBH data only (left) and incorporating SBH data (right) for the time periods 1980--1985 and 2010--2015. Denser hatching reflects greater uncertainty (based on the posterior standard deviation of logit U5MR).}
\label{fig:q5mapa}
\end{figure}

\begin{figure}[tbp]
\centering
\includegraphics[height=5in,width=7in
]{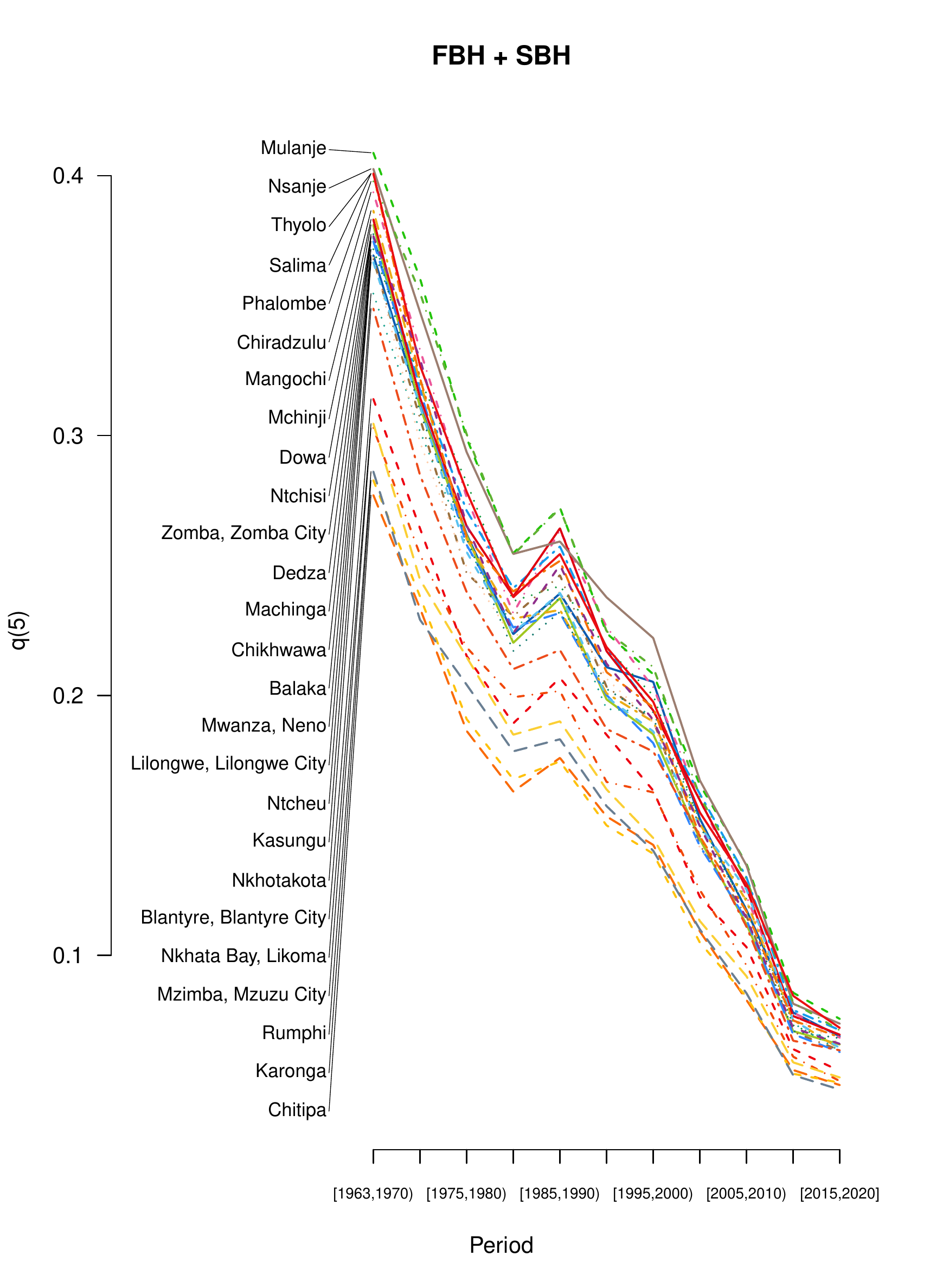}
\caption{Trends of U5MR over time for the 26 regions in Malawi.}
\label{fig:q5spaghetti}
\end{figure}

Figure \ref{fig:q5spaghetti} provides the trends in U5MR (posterior median) over time for the 26 regions,  from the FBH + SBH model. The figure shows a decrease in all regions until the 80s, some leveling, and then a decrease.
There is little crossing over of the curves, which shows that the interaction between space and time  is not a significant contribution. 
In the SBH + FBH model for the time period 2010--2015 the regions with the lowest U5MR (posterior median less than 56 per 1000) were Rumphi, Karonga, and Chitipa, the three northern most regions. The regions with highest U5MR (posterior median greater than 81 per 1000) were Nsanje, Phalombe, Thyolo, and Mulanje, which are the four regions in the south east.  Overall, we see that the absolute subnational variation in U5MR has decreased over time but in relative terms there is still significant subnational variation. 
Specifically, the overall range for 2010--2015 was 54 (Rumphi) to 86 (Mulanje). For the period 1970--1975 the range was 229 (Rumphi) to 360 (Mulanje). So although there has been a dramatic decline, the relative range of (largest-smallest)/smallest is 0.57 in both periods, i.e,~the area with the highest U5MR is 57\% greater than the lowest. This level of spatial detail is much greater than is available in the DHS and MICS reports.

\clearpage

\section{Discussion}

In this paper we have described a flexible model  for U5MR estimation that combines FBH and SBH data, in an efficient implementation.  Code to implement the methods is available at \url{http://faculty.washington.edu/jonno/software.html}. The Malawi analysis took 7 minutes to run on a laptop. 
Since we have specific models for both fertility and mortality, it is straightforward to include spatial and temporal smoothing models for each, and of a type that the context requires (or the user is familiar with). We used random walk and intrinsic CAR models, but it is straightforward to use other forms, such as temporal spline models \citep{alkema:new:14} or the Leroux spatial model \citep{riebler:etal:16}. We focused on U5MR, but
 mortality estimates for ages under 5 are natural (and easy) to obtain, given our discrete hazards mortality model. We could also include covariates in our models, for example, mother's education has been found to be associated with U5MR \citep{golding:etal:17}.
 
We described the model  using mother's age but the approach could be extended, if time since first birth or time since marriage \citep{manualX:83,hill1999child} were deemed preferable time scales. Currently, the United Nations produce national estimates of U5MR that incorporate SBH data using indirect estimates obtained from the time since first birth variant \citep{alkema2014child,hill:etal:12}. 

A major difficulty that has long plagued the synthesis of FBH and  SBH data is reconciling systematic differences between the two data sources. In work by \cite{wilson:wakefield:20SBH} they find systematic differences in data from Malawi, and differences have also been found in other applications \citep{hill:etal:15}. In our example and in \cite{wilson:wakefield:20SBH} we have attributed the differences to biases in the SBH data, but
 biases also exist in FBH data from household surveys, and some surveys are not used because they are thought to be unreliable. For examples, see the Supplementary Materials of \cite{li:etal:19} in which there is a list of surveys that were not used in estimation of the U5MR in sub-Saharan Africa. However, where local and contextual information are available, the model we have proposed allows the flexible modeling of  biases in either data source in a transparent manner. 
Theoretically, we could analyze SBH data alone, if we had relevant fertilities to use in the model, but this endeavor is inherently dangerous, due to the close to non-identifiability of the SBH data alone.

In conclusion, while there are many challenges to modeling SBH data, the Brass method  has proved very useful over the many years since its introduction, and the modeling approach we describe provides the potential to leverage SBH data to an even greater extent.

\clearpage
\begin{center}
\Large{\bf Supplementary Materials for ``A Probabilistic Model for Analyzing Summary Birth History"} 

\end{center}

\section*{Appendix}

\subsection*{Appendix: Further Details on the Brass Method}

\begin{table}[!h]
\centering
\begin{tabular}{r|cc}
Age group & $i$ & $x$ \\
\hline
15--19 & 1 & 1 \\
20--24 & 2 & 2 \\
25--29 & 3 & 3 \\
30--34 & 4 & 5 \\
35--39 & 5 & 10 \\
40--44 & 6 & 15 \\
45--49 & 7 & 20 
\end{tabular}
\caption{The correspondence between mothers age group and mortality age which is used in the Brass method. The probability of death before age $x$, is defined as $q(x)$. The mother's age group is indexed by $i=1,\dots,7$, and $x$ is the corresponding age of children for whom cumulative mortality is best identified by  the proportion of children who died to mothers in age group $i$. So, for example, the proportion of children who died to mothers in the age range 30--34 ($i=4)$ estimates $q(5)$, the U5MR. }
\end{table}

\section*{Appendix: Further Details on The Full Model}

%
Here we extend the simplified derivation described in ``An Intuitive Derivation of the New Method'' to the more realistic scenario in which the births and mortality models are more complex.
Let $f_m(\bx_t)$ denote the probability a woman gives birth at age $m$ and in year $t$ with $\bx_t$ containing the covariates at time $t$ associated with birth. For mortality, we use a discrete hazards model. Let $_1 q_a(\bx_t) = q_a(1,\bx_t)$ denote the risk of mortality, i.e.,~the probability that a child dies between age $a$ and $a+1$ with $\bx_t$ now containing the covariates at time $t$ associated with mortality. Although we use $\bx_t$ for covariates in both models for notational convenience, the covariates used for each will generally differ. The parameter of interest is $q(5,\bx_t)$, the probability of death within 5 years of life, \textit{at time} $t$ and with covariates $\bx_t$. Here,
\begin{align}
q(5,\bx_t) = 1 - \prod_{i=0}^{4}(1 -\, _1q_i(\bx_t)). \label{eq:trueq5sup}
\end{align}
 Let $q^\star(a, \bx_t)$ be the probability of dying within $a$ years \textit{given birth at time} $t$ and with covariates $\bx_t$; thus,
 \begin{align}
 q^\star(a,\bx_t) = 1 - \prod_{i=0}^{a-1}(1 -\, _1q_i(\bx_{t+i})). \label{eq:starq5sup}
 \end{align}
The difference between Eqs.~(\ref{eq:trueq5sup}) and (\ref{eq:starq5sup}) for $a=5$ is subtle, but crucial. In Eq.~(\ref{eq:trueq5}), we envisage a synthetic cohort of children that are born in year $t$ and then repeat year $t$  five times, meaning they experience in the same calendar year the mortalities of each age band. In  Eq.~(\ref{eq:starq5sup}), we instead imagine a real cohort of children that are born in year $t$ and are followed up to year $t+5$.

For FBH data, where information is available on when births and deaths occurred, let  $Y_{mt}$ be an indicator for birth in year $t$ to a woman of age $m$ years (in the case of multiple births, we include multiple indicators). Let $Z_{at}$ be an indicator for death between ages $a$ and $a+1$  in years $t$ to $t+1$. To summarize, a reasonable model for the FBH data consists of the birth and death components:
\begin{align*}
Y_{mt} | f(m,\bx_t) & \sim \text{Bernoulli}(f_m(\bx_t)), \\
Z_{at} | _1q_a(\bx_t) & \sim \text{Bernoulli}( _1q_a(\bx_t)). 
\end{align*}

Now consider SBH data from a survey taken in year $t_s$. For women who are of age $m_s$ years at the time of the survey, define $T_B^{m_s}(\bx_t)$ to be the total number of children ever born to  those women who have covariates $\bx_t$ for all $t$. Further, for these women define the (unobserved) number of children born $a$ years prior to the survey as $B_a(\bx_{t_s-a})$ . Note that $\sum_{a=0}^{m_s}B_a(\bx_{t_s-a}) = T_B^{m_s}(\bx_t)$. Similarly, define $T_D^{m_s}(\bx_t)$ to be the total number of children that ever died to women who are $m_s$ at the time of the survey who have covariates $\bx_t$ for all $t$. Define the (unobserved) number of children that were born $a$ years prior to the survey \textit{and} died by the time of the survey (i.e.,~died within $a$ years), to be $D_a( \bx_{t_s - a})$. Again, $\sum_{a=0}^{m_s}D_a(\bx_{t_s - a}) = T_D^{m_s}(\bx_t)$. Therefore, a reasonable model for the unobserved data is,
\begin{align*}
D_a(\bx_t) | B_a(\bx_t), q^\star(a,\bx_t) \sim \text{Binomial}(B_a(\bx_t), q^\star(a,\bx_t)),
\end{align*}
which provides a starting point for deriving an approximation to the distribution of the SBH data.
Approximating the Binomial with a Poisson,
\begin{align*}
D_a(\bx_t) | B_a(\bx_t), q^\star(a,\bx_t) \sim \text{Poisson}(B_a(\bx_t)q^\star(a,\bx_t)).
\end{align*}
Suppressing the dependence on $\bx_t$, for notational convenience, and summing over age, gives,
\begin{align}
T_D^{m_s}| \bB^{m_s}, q^\star(a) \sim \text{Poisson}\left(\sum_{a=0}^{m_s}B_aq^\star(a)\right). \label{eq:sbhtrue}
\end{align}
Finally, we sum over all $(m_s+1)^{T_B^{m_s}}$ possible combinations of when births could have occurred for a given $T_B^{m_s}$ to obtain a mixture distribution. We let $S_{m_s}$ represent the set of all legal configurations of births that can sum to $T_B^{m_s}$. Then,
\begin{align}
T_D^{m_s} |  T_B^{m_s}, \boldf_m, \bq^\star(a)  \sim \sum_{\bB \in S_{m_s}}  \Pr(\bB^{m_s} | T_B^{m_s}, \boldf_m) 
\times  \text{Poisson}\left(\sum_{a=0}^{m_s}B_a q^\star(a)\right) , \label{eq:sbhmix}
\end{align}
where $\boldf_m$ are the birth probabilities for a woman of age $m$.
%

We approximate Eq.~(\ref{eq:sbhmix}) by
\begin{align}
T_D^{m_s} |  T_B^{m_s}, \bc, \bq^\star(a) & \sim  \text{Poisson}\left(T_B^{m_s}\sum_{a=0}^{m_s}c_{m_s}(a) q^\star(a)\right), \label{eq:sbhsimplified}
\end{align}
where
$$
c_{m_s}(a,\bx_{t_s-a})  = \frac{f_{m_s -a}(\bx_{t_s-a})}{\sum_{a=0}^{m_s}f_{m_s -a}( \bx_{t_s-a})}. \nonumber
$$
Since $\bc$ is often unknown, we propose first fitting a fertility model to the FBH data. Such a model allows an estimate of $f(m,\bx_t)$ to be formed, i.e.,~$\widehat{f}(m,\bx_t)$, which can then be transformed to $\widehat{c}_{m_s}(a,\bx_t)$. In the simulation study that we next describe, we  investigate the effect of both replacing Eq.~(\ref{eq:sbhmix}) with  Eq.~(\ref{eq:sbhsimplified}) and compare with estimating $\bc$.


\section*{Appendix: Simulation Details and Further Results}

Birth histories for $47 \times 24,000$ women were simulated on a discrete, yearly time scale. For simplicity, we allowed the year prior to the survey to be completely observed and did not allow for births during the survey year, which follows the simple example provided in Section 2.1. Thus, children could be born at any point prior to and including $t_s -1$ (when the woman was aged $m_s -1$), and could die in $t_s$ where $t_s$ is the year of the survey and $m_s$ is the age of the woman at time of survey.

\begin{figure}[tbp]
  \centerline{\includegraphics[width=5in]{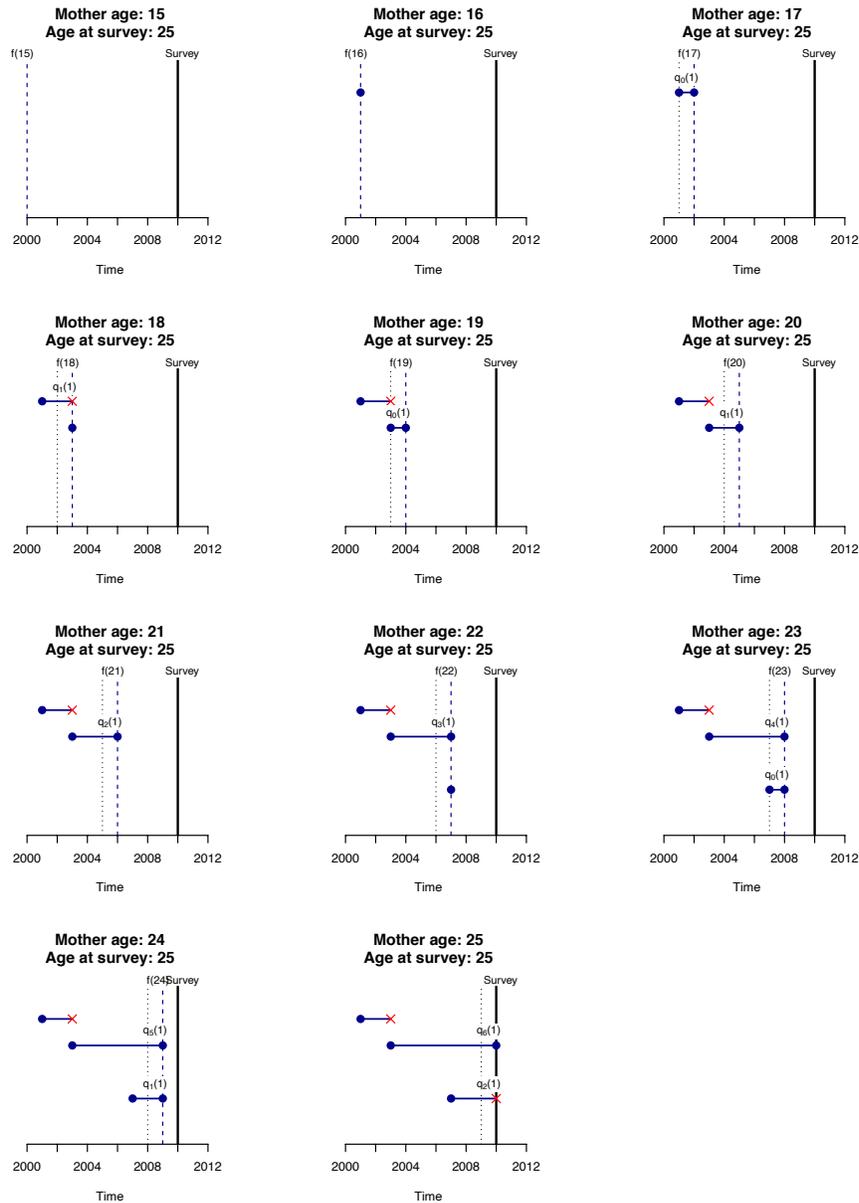}}
  \caption[Illustration of the data generating mechanism along with relevant probabilities]{Illustration of the data generating mechanism along with relevant probabilities. Suppose a woman is 25 at the time of the survey in 2010 and suppose $f(m)>0$ for $m\geq15$. Starting at the top right and proceeding left and down are panels following her and any children she has forward through time starting at age 15. The blue dashed line represents the current year and black dotted line represents the prior year. Blue circles represent births and survival, red ``x''s represent deaths.}
\vspace*{-3pt}
\label{Fig:datagen}
\end{figure}

Figure \ref{Fig:datagen} 
illustrates how FBH data was simulated for a woman who was $25$ at the time of the survey. In the top left panel, when the woman is $15$, the probability she gives birth is $f(15)$ (fertility does not change over time). In this example, she does not give birth. In the middle top panel, when the woman is $16$, the probability she gives birth is $f(16)$. Here, she does give birth. In the following year (top right panel), the probability the woman gives birth is $f(17)$, and the probability the child dies is $_1q_0 = q_0(1)$. As the woman and her children age, we observe her to have three children at ages 16, 18, and 23. One child dies between age 1 and 2 and another dies between age 2 and 3. Her other child survives through the time of the survey. For women with FBH data, this information is completely observed. For women with SBH data, we only observe the total number of children the woman had and the number of those children that died (in this example, three births and two deaths).

\begin{figure}[!h]
\centering
\includegraphics[width=0.75\linewidth]{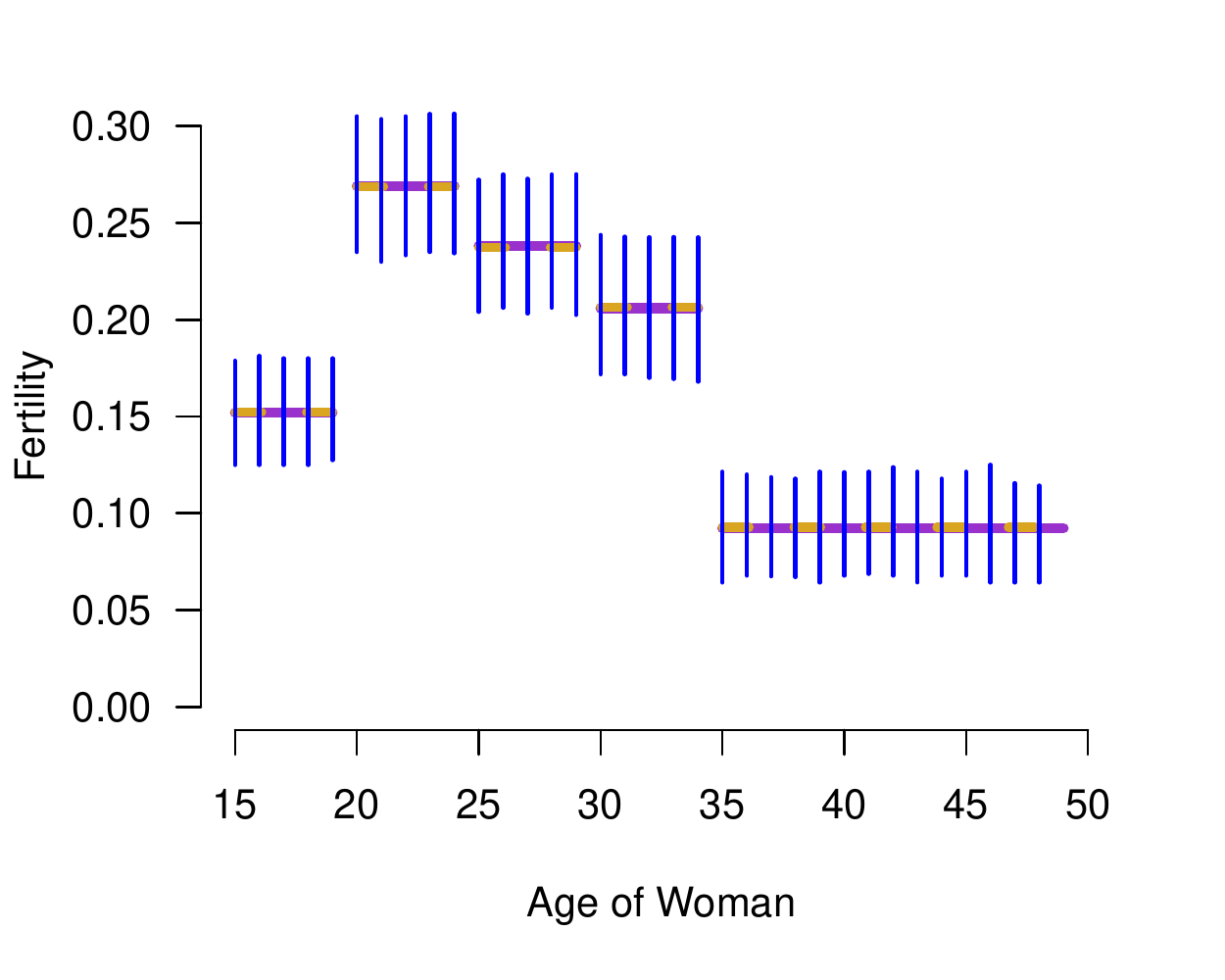}
\caption{Yearly fertility estimates under different models. Blue: 5th and 95th percentiles of observed fertility probabilities in the SBH data (by region and surveyed woman's age). Purple: true underlying fertility probabilities. Yellow: estimated fertility probabilities from FBH data.}
\label{fig:fert-sim}
\end{figure}

Figure \ref{fig:fert-sim} shows the yearly fertilities that were used and the estimates.

The estimates of U5MR, and corresponding measures of uncertainty, by region and time period and corresponding measures of uncertainty by region and time period, were derived using a multivariate normal approximation. That is, defining $\widehat{\bpsi}$ to be the estimates for a generic region and time period and $\widehat{\bSigma}$ to be the inverse Hessian obtained from using \texttt{TMB}, we simulate 1,000 draws, from $\widehat{\bpsi}^{(i)} \sim N\left(\widehat{\bpsi}, \widehat{\bSigma}\right),$

\begin{figure}[tbp]
    \centering
    \includegraphics[width=\linewidth]{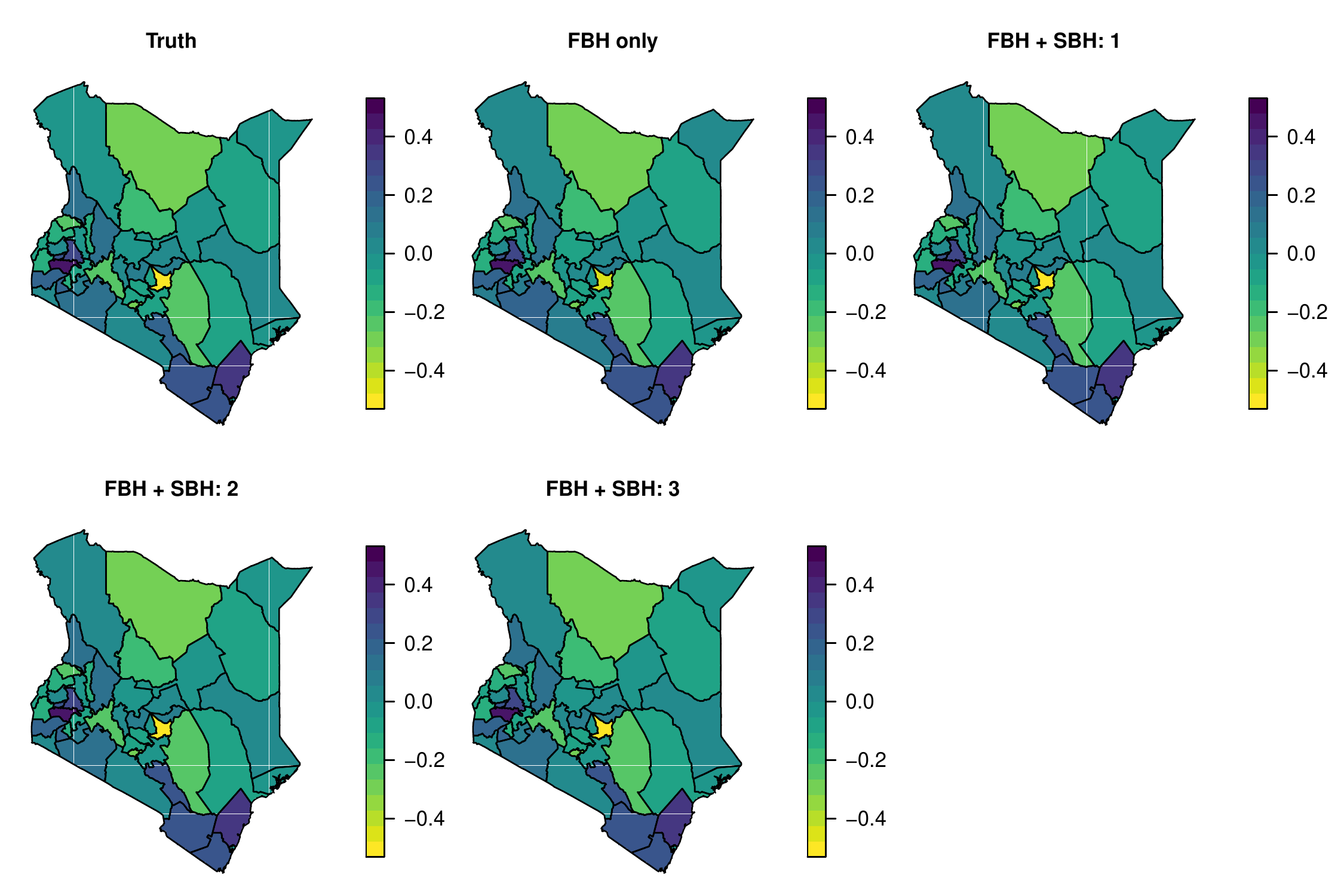}
    \caption[Mortality structured and unstructured spatial random effect parameters in 
    ]{Structured and unstructured spatial random effect parameters, $S_r + \epsilon_r$, in mortality model. Posterior median estimates are mapped. The truth is compared to FBH analysis and three approaches to dealing with births in the SBH model:  (1) true births, (2) true fertilities, (3) estimated fertilities.}
    \label{fig:bv:spatsim}
\end{figure}

\clearpage
\section*{Appendix: Further Information for Malawi Application}

\begin{figure}[!h]
\centering
\includegraphics[width=0.9\linewidth]{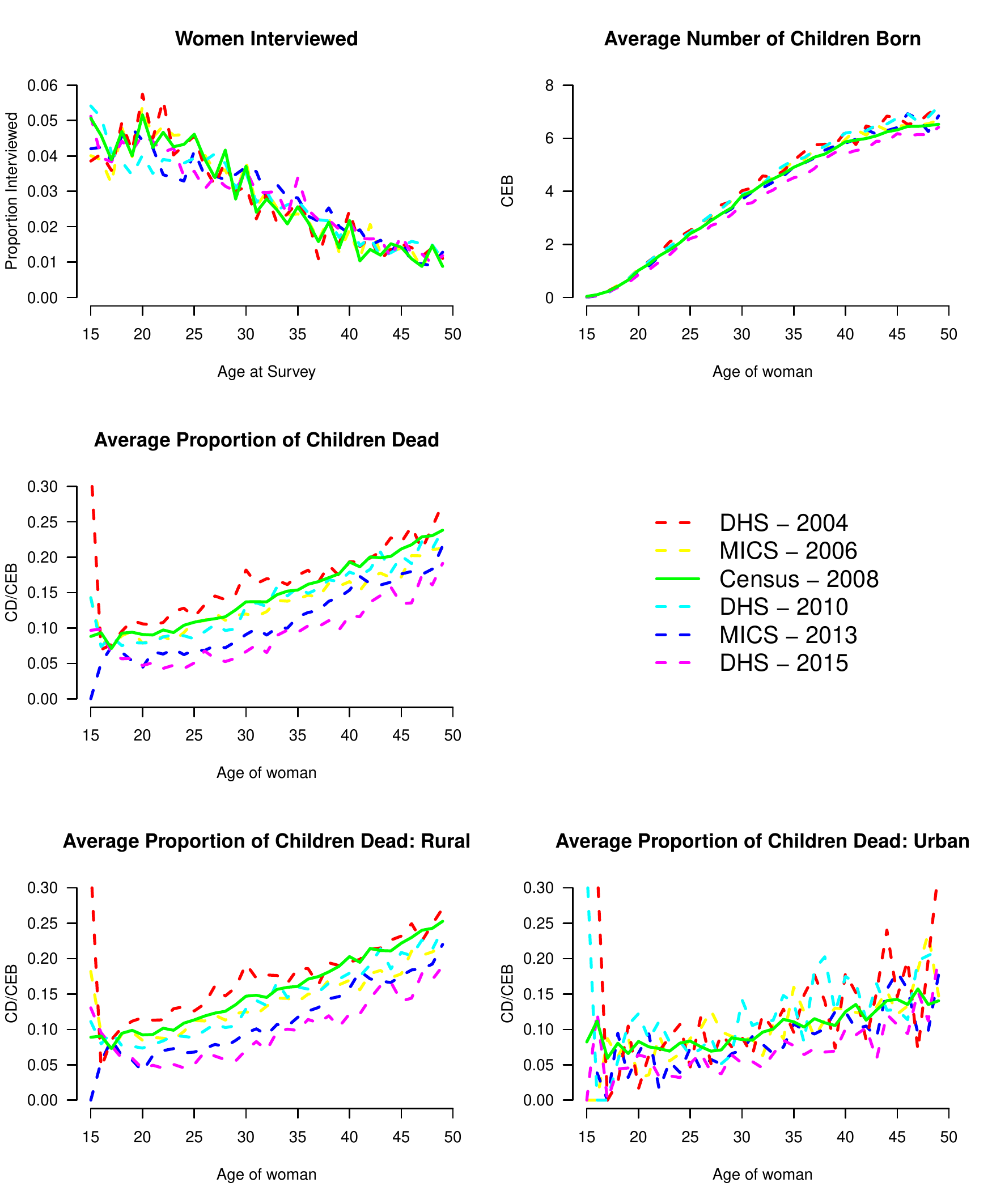}
\caption{Exploratory plots showing proportion of women interviewed, average number of reported children born, and average proportion of children dead (overall and by strata) by age of woman at the interview. Colors indicate the survey or census year and type.}
\label{Fig:ap_eda}
\end{figure}

We conducted an exploratory data analysis, comparing the five FBH surveys and census. Results are in Figure \ref{Fig:ap_eda}. Across all data sources, the age of woman at interview tends to be similar. The reported total number of children born tends to be smaller for more recent surveys, consistent with decreasing fertility over time. This also tends to be the case for the average proportion of children died, consistent with decreasing mortality. However, we do notice that for rural women in the census, they tended to report a higher proportion of their children died than we would expect given the other surveys.

\begin{figure}[!h]
\centering
\includegraphics[width=0.45\linewidth]{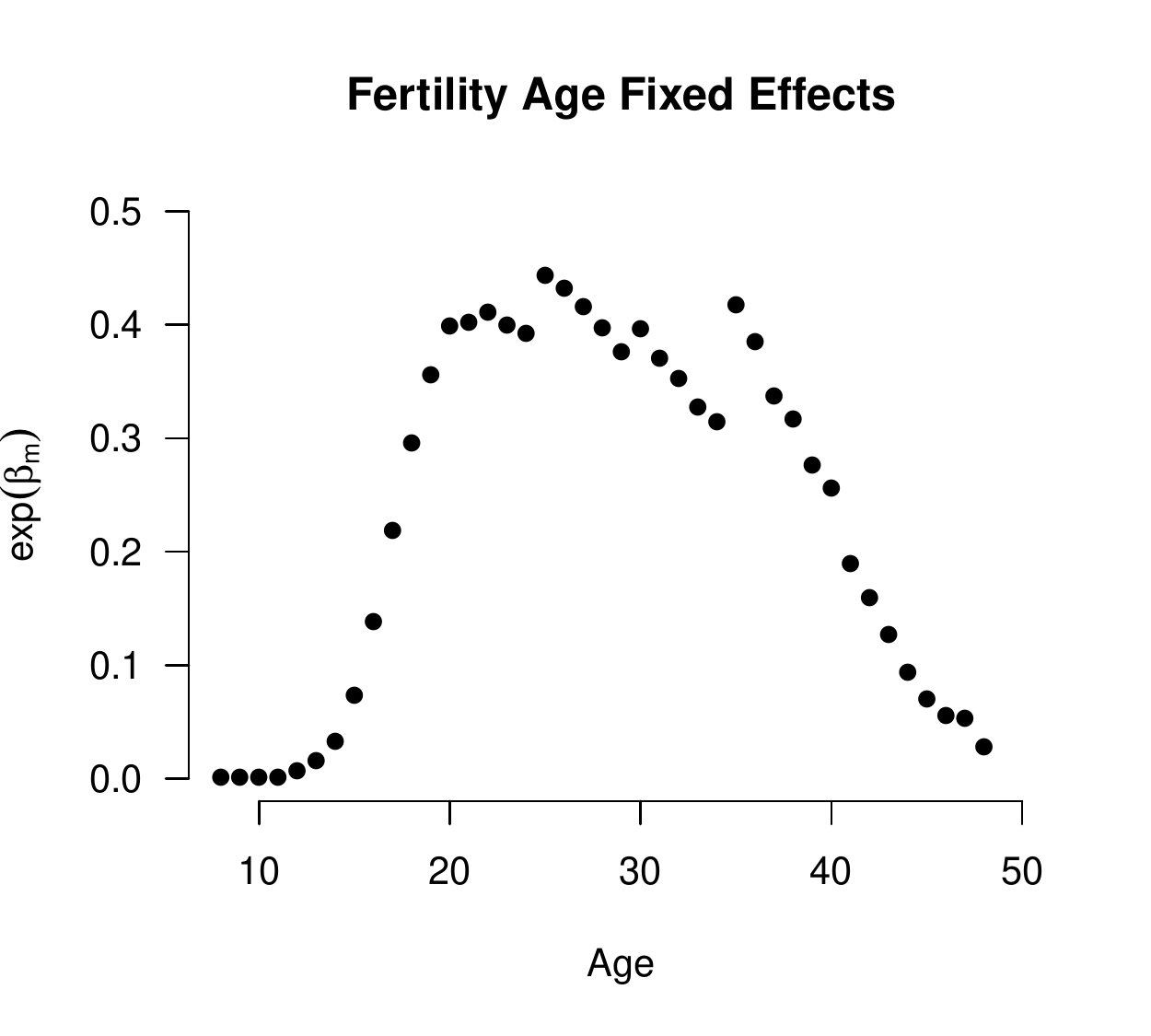}
\includegraphics[width=0.45\linewidth]{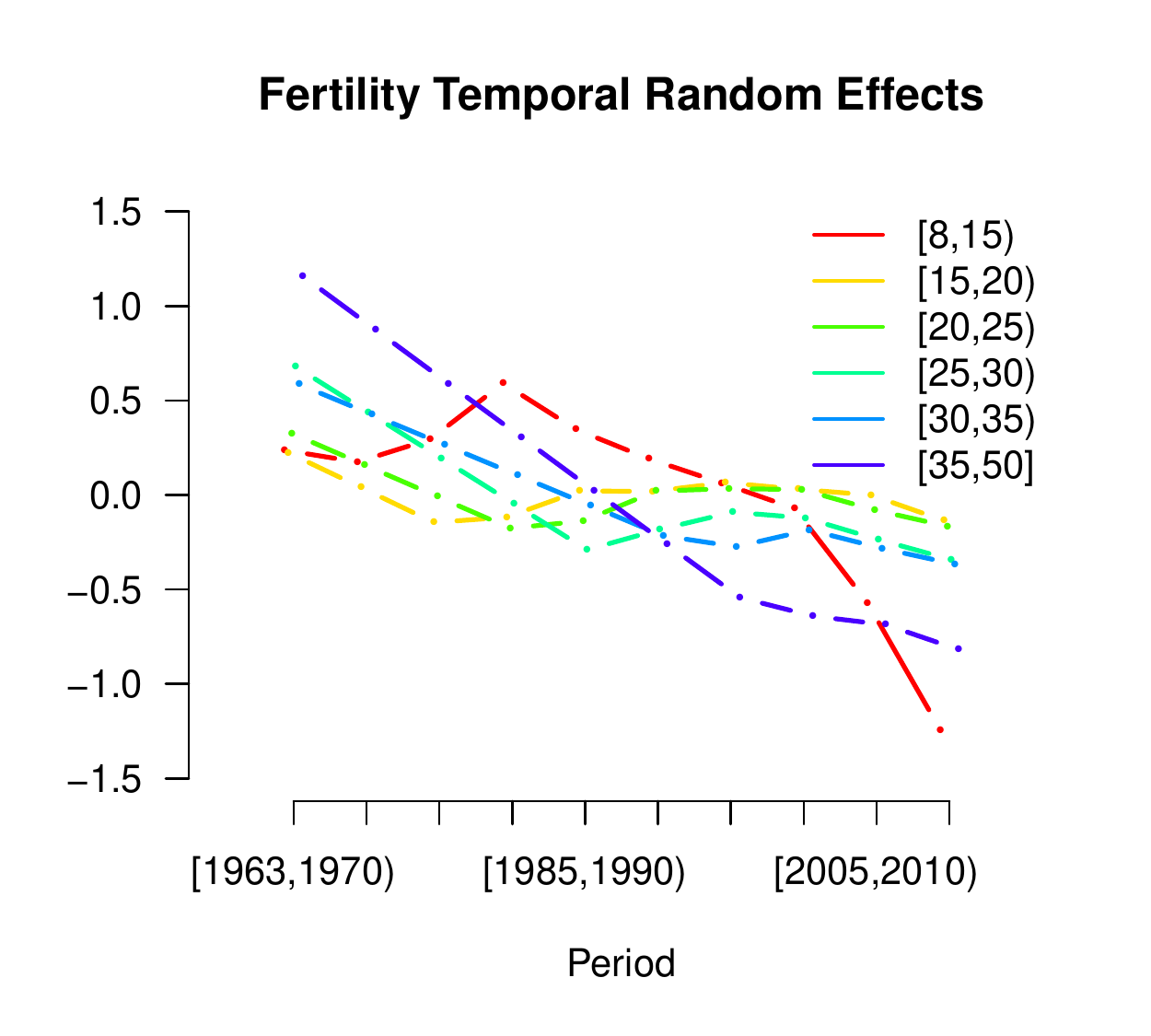}
\caption{Left: Posterior medians of fertility intercepts by age $\exp(\beta_m)$. Right: Posterior medians of time trends $\phi_{c[m]}(p)$}
\label{Fig:mfert-age}
\end{figure}

The parameter estimates for the fertility model are:
\begin{itemize}
\item $\beta_{\text{URB}}$: -0.219 (-0.226, -0.198)
\item $\kappa_T$: 20.2 (11.3, 35.2)
\item $\kappa_S$: 367 (117, 1620)
\item $\kappa_\epsilon$: 517 (163, 1280) 
\end{itemize}
Figure \ref{Fig:mfert-age} shows estimates of the fertility odds (left) and the time trends (right).

\begin{figure}[!h]
\centering
\includegraphics[width=0.45\linewidth]{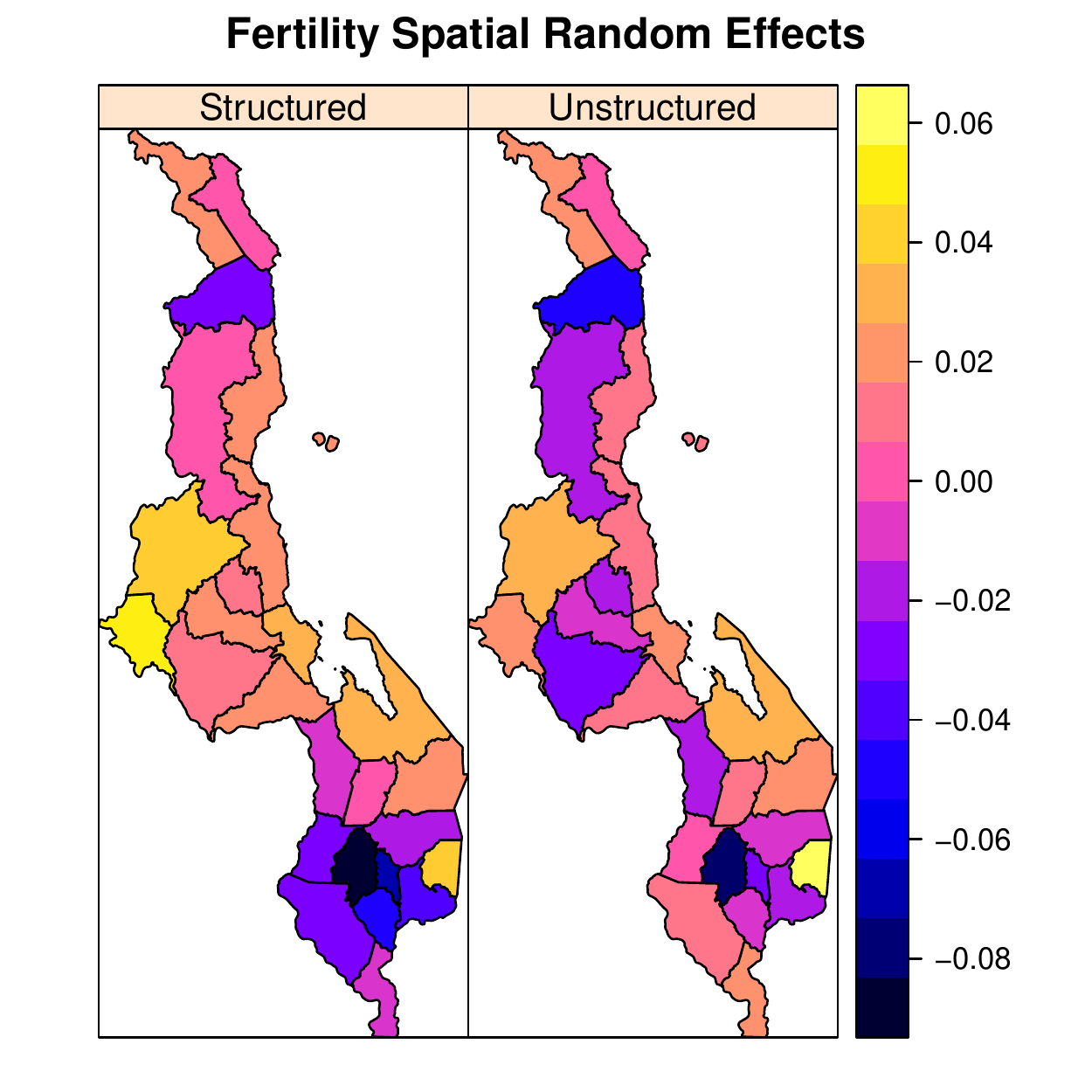}
\includegraphics[width=0.45\linewidth]{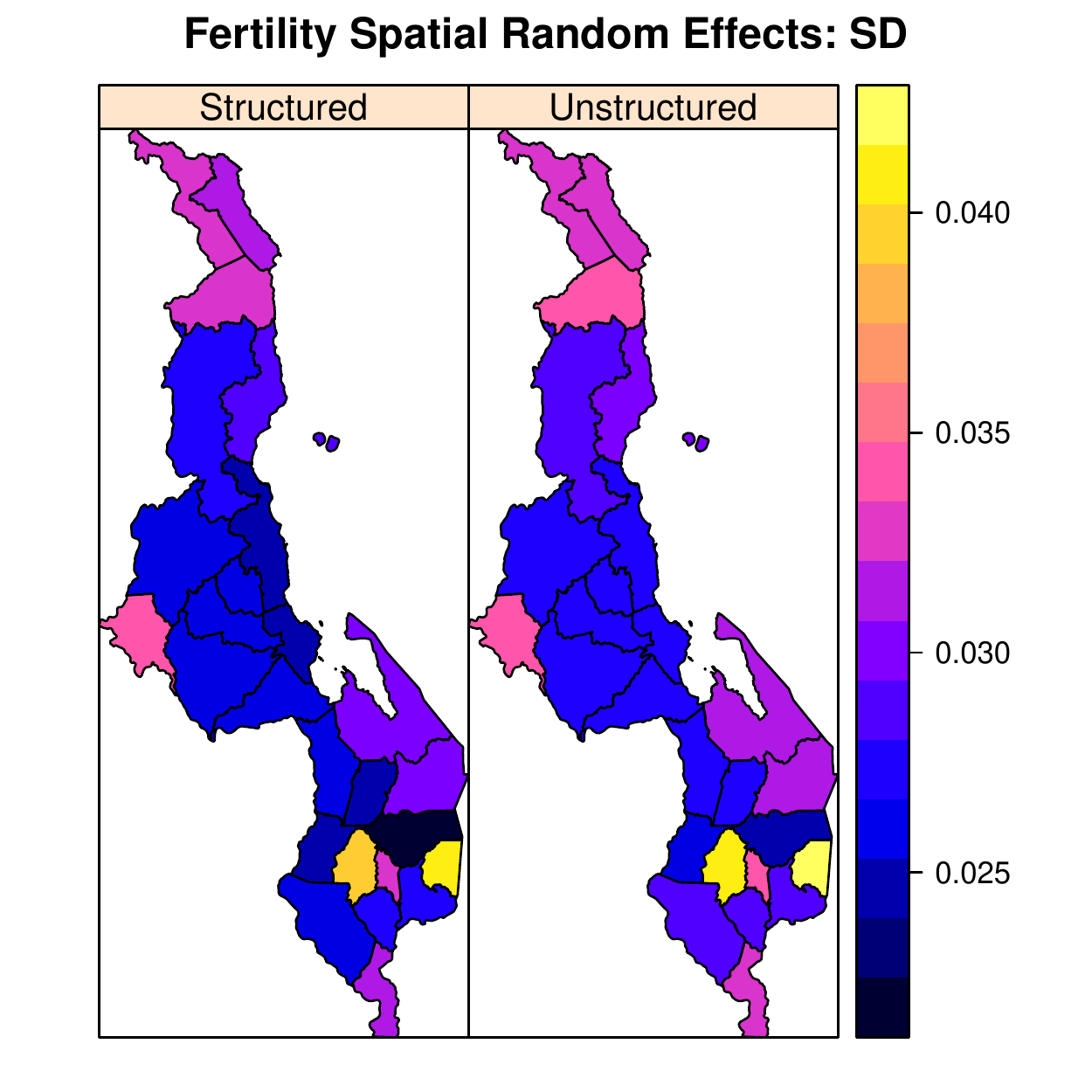}
\caption{Malawi results for the regional structured $S_r$ and unstructured  $\epsilon_r$ random effects for fertility. Left: Posterior medians. Right: Posterior standard deviations.}
\label{Fig:mfert-sre}
\end{figure}

The region-specific spatial and iid adjustments $S_r$, $\epsilon_r$ for fertility are shown in Figure \ref{Fig:mfert-sre}; on the left we give point estimates, and on the right measures of uncertainty. It is clear that the between region variation is mostly spatially structured, with only a small contribution from the random shocks. The uncertainty in the spatial random effects is greater also.
 
\begin{figure}[!h]
\centering
\includegraphics[width=0.75\linewidth]{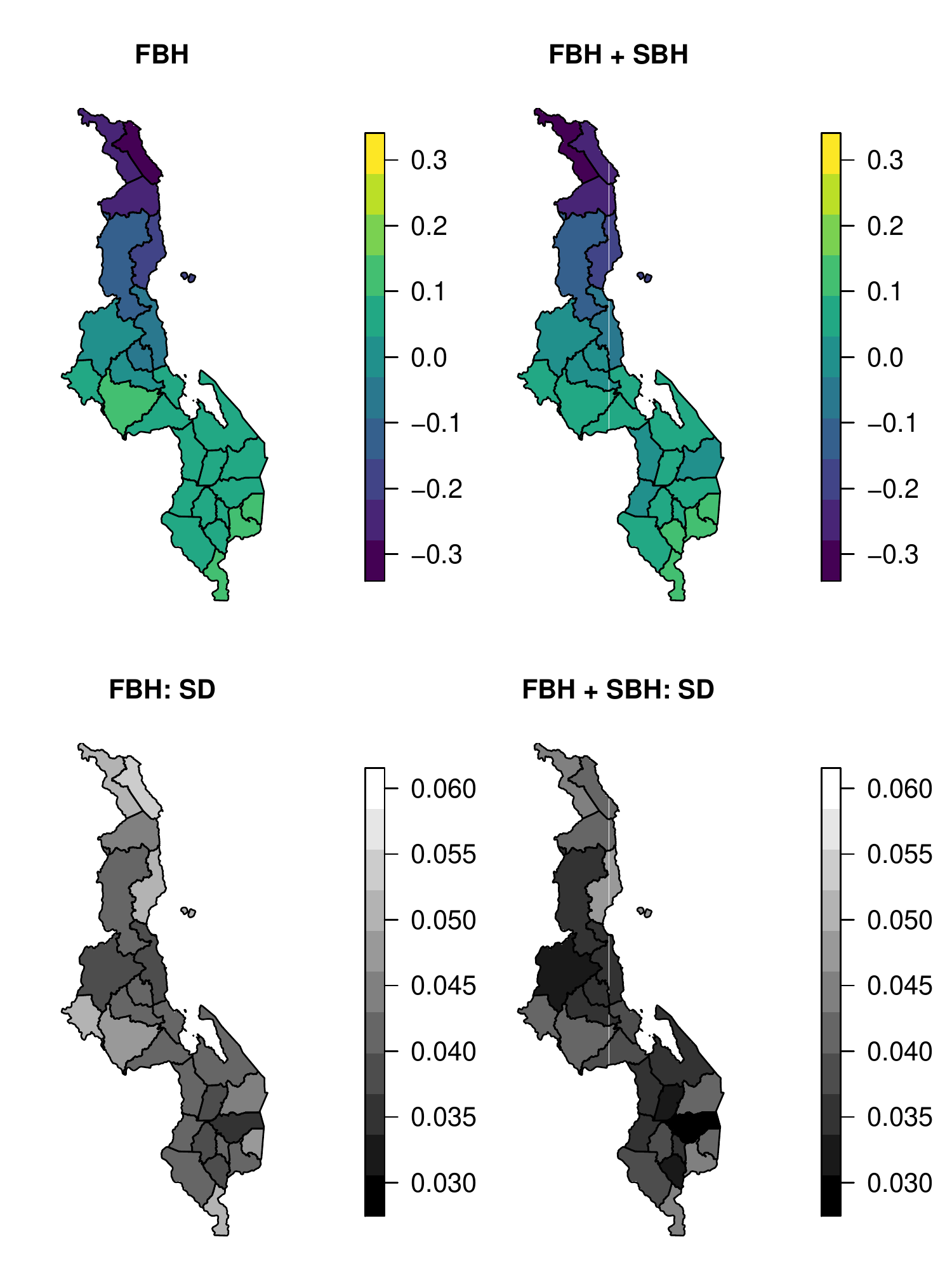}
\caption{Spatial random effect, $S_r$, comparison of FBH (left) and  FBH + SBH (right) analyses. Top: posterior medians. Bottom: posterior standard deviations.}
\label{fig:malmorspat}
\end{figure}

\begin{figure}[!h]
\centering
\includegraphics[width=0.75\linewidth]{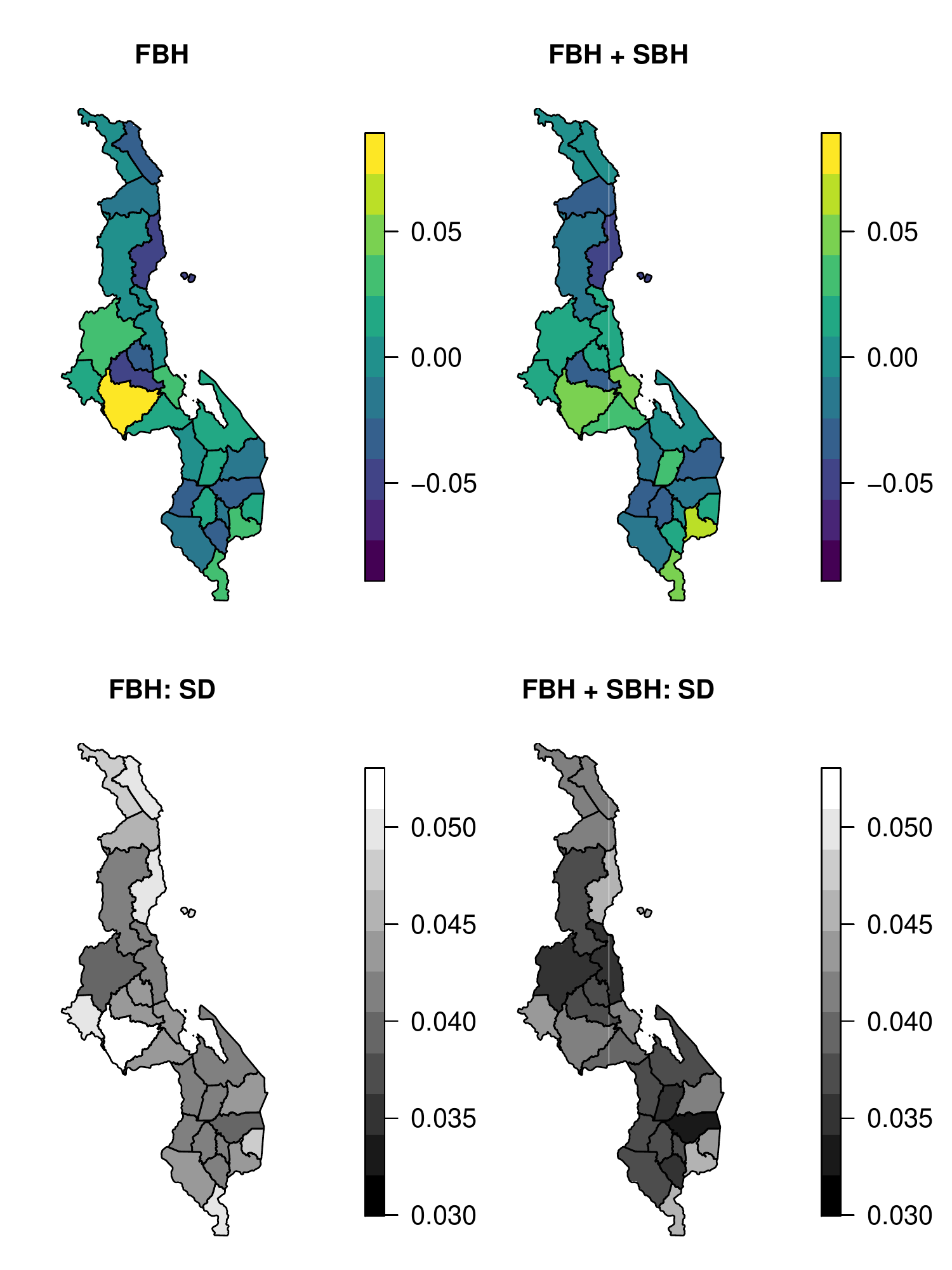}
\caption{Unstructured random effect, $\epsilon_r$, comparison of FBH (left) and  FBH + SBH (right) analyses. Top: posterior medians. Bottom: posterior standard deviations.}
\label{fig:malmorspateps}
\end{figure}

Figures \ref{fig:malmorspat} and \ref{fig:malmorspateps} show point estimates of the spatial and iid regional terms, respectively, along with measures of uncertainty and shows that, as with fertility, there is strong spatial structure. Specifically, there is an increasing trend in mortality, when moving from north to south.

\clearpage
\section*{Appendix: Computation}

This model is implemented in the \texttt{TMB} \texttt{R} package. The \texttt{TMB} package has a less user-friendly interface than \texttt{R-INLA}, but has far fewer restrictions on the class of models that can the fit. In particular, \texttt{R-INLA} cannot be used for our model because of the nonlinear mean function. Let $\bz$ denote the vector containing $z_a(t)$, $\mathbf{d}$ the vector containing the realized values of $D_{m_s}$, $\mathbf{B}$ the vector containing $B_{m_s}$, $\bc$ the vector containing $c_{m_s}(a, \bx(t))$, $\bbeta$ the vector containing latent mortality fixed effects, and $\btheta$ the vector containing latent mortality random effects. In our proposed approach, the negative log posterior is
\begin{align*}
-f(\btheta, \bbeta) = \underbrace{-\log p(\bz|\bbeta,\btheta)}_{\text{FBH contribution}} - \underbrace{\log p(\mathbf{d}|\mathbf{B}, \bc, \bbeta,\btheta)}_{\text{SBH contribution}} - \underbrace{\log p(\bbeta, \btheta)}_{\text{prior}}.
\end{align*}
This is specified in a C++ template.
After specifying the objective function, the user calls a \texttt{TMB} function that compiles the code and the user can flag any parameters as random effects, i.e., $\btheta$. \texttt{TMB} uses Laplace approximations to integrate out the random effects. Specifically,
\begin{align*}
L(\bbeta) &= \int \exp\left[-f(\btheta, \bbeta)\right]d\btheta\\
&\approx  L^\star(\bbeta)= \text{det}\{H(\bbeta)\}^{-1/2}\exp\left[-f(\hat{\btheta}(\bbeta),\bbeta )\right],\\
\end{align*}
where $H(\bbeta) = - \frac{\partial^2}{\partial \btheta^2} f(\btheta,\bbeta) |_{\btheta = \hat{\btheta}(\bbeta)}$ and $\hat{\btheta}(\bbeta) = \text{argmin}_{\btheta} f(\btheta, \bbeta)$. The \texttt{TMB} function returns $-\log L^\star(\bbeta)$ and its derivative so that an estimate for $\bbeta$ can be obtained using nonlinear optimization techniques,
\begin{align*}
\hat{\bbeta} & = \text{argmin}_{\bbeta} - \log L^\star(\bbeta),
\end{align*}
and the Hessian can be used to derive an estimate of the uncertainty.



Use the \texttt{TMB} package
\begin{itemize}
\item To get constrained parameters in the \texttt{TMB} algorithm use \nocite{rue:knorrheld:05} Rue and Held (2005, Algorithm 2.30):
	\begin{itemize}
	\item The ``parameters'' are unconstrained (define these as $U(r)$ and $\phi_{c[a]}(t)$) and then the algorithm is applied to get the constrained versions
	\item First define $Q_j = \kappa_j K_j + 10^6 
	$ for $j=\{T,S\}$ 
	\item In algorithm 2.30, $\bA_j = [1,\dots,1]$ where the length of $\bA_t$ is the number of periods and $\bA_s$ is the number of regions
	\end{itemize}
\end{itemize}

\section*{Appendix: Prior Specifications}

Some care is in general required when specifying prior distributions, particularly for variance components. A very appealing and rigorous approach has been developed recently \citep{simpson:etal:17}, under the name { penalized complexity} (PC) priors. The basic idea is to take a baseline (simple) model and then penalize departures from this model. For example, it is well-known that there can be sensitivity to the prior on the variance, when random effects models are fitted to data. Under the PC model, the baseline model corresponds to a variance of zero (in which case all random effects are zero), and greater values of the random effects standard deviation are more and more penalized (i.e.,~discouraged in the prior). To specify these priors, one sets two values for each parameter, a value of the parameter (on an interpretable scale), below we call this $u$,  and a prior probability of exceedance of this value, which we call $\alpha$. 

For the mortality model in the simulation we use the following as (independent) priors,
\begin{align*}
\beta_{c} \sim N(0, 10^2), \qquad & \kappa_T \sim \text{PCprior}(u=1,\alpha=0.01),\\
\kappa_S \sim \text{PCprior}(u=1,\alpha=0.01), \qquad & \kappa_\epsilon \sim \Gamma(\text{shape}=1,\text{scale}=200).
\end{align*}

For our Malawi model, we specify the following priors. For fertility:
	\begin{itemize}
	\item  $\kappa_T \sim \text{PCPrior}(u=0.5, \alpha = 0.01)$ (precision for $\phi_{c[m]}(p)$, though $u$ is on the standard deviation scale)
	\item $\kappa_S \sim \text{PCPrior}(u=0.5, \alpha = 0.01)$ (precision for $S_r$, though $u$ is on the standard deviation scale)
	\item $\kappa_\epsilon \sim \Gamma(1, 1/400)$ (precision for $\epsilon_r$)
	\item $\beta_X \sim N(0,100)$
	\end{itemize}
	For mortality:
	Priors:
	\begin{itemize}
	\item  $\kappa_T \sim \text{PCPrior}(u=1, \alpha = 0.01)$ (precision for $\phi_{c[a]}(p)$)
	\item $\kappa_S \sim \text{PCPrior}(u=1, \alpha = 0.01)$ (precision for $S_r$)
	\item $\kappa_\epsilon \sim \Gamma(1, 1/200)$ (precision for $\epsilon_r$)
		\item $\beta_X \sim N(0,100)$
		\item $\beta_{\text{SBH}}, \beta_{\text{SBH,URB}} \sim N(0, 10)$
		
	\end{itemize}

\clearpage
\bibliography{spatepi}

\end{document}